\documentclass[12pt]{article}
\usepackage[mathscr]{eucal}
\usepackage{amssymb}
\input epsf
\makeatletter

\def\newpic#1{}

\voffset= - 0.5cm

\def\hybrid{\topmargin 0pt      \oddsidemargin 0pt
             \headheight 0pt \headsep 0pt

       \textheight 9in         
             \textwidth 6.25in       
             \textheight 9.5in       
             \marginparwidth 0.0in
             \parskip 5pt plus 1pt   \jot = 1.5ex}
\catcode`\@=11
\def\marginnote#1{}

\newcount\hour
\newcount\minute
\newtoks\amorpm
\hour=\time\divide\hour by60
\minute=\time{\multiply\hour by60 \global\advance\minute by-\hour}
\edef\standardtime{{\ifnum\hour<12 \global\amorpm={am}%
             \else\global\amorpm={pm}\advance\hour by-12 \fi
             \ifnum\hour=0 \hour=12 \fi
             \number\hour:\ifnum\minute<10 0\fi\number\minute\the\amorpm}}
\edef\militarytime{\number\hour:\ifnum\minute<10 0\fi\number\minute}

\def\draftlabel#1{{\@bsphack\if@filesw {\let\thepage\relax
        \xdef\@gtempa{\write\@auxout{\string
           \newlabel{#1}{{\@currentlabel}{\thepage}}}}}\@gtempa
        \if@nobreak \ifvmode\nobreak\fi\fi\fi\@esphack}
             \gdef\@eqnlabel{#1}}
\def\@eqnlabel{}
\def\@vacuum{}
\def\draftmarginnote#1{\marginpar{\raggedright\scriptsize\tt#1}}

\def\draftlabel#1{{\@bsphack\if@filesw {\let\thepage\relax
        \xdef\@gtempa{\write\@auxout{\string
           \newlabel{#1}{{\@currentlabel}{\thepage}}}}}\@gtempa
        \if@nobreak \ifvmode\nobreak\fi\fi\fi\@esphack}
             \gdef\@eqnlabel{#1}}
\def\@eqnlabel{}
\def\@vacuum{}
\def\draftmarginnote#1{\marginpar{\raggedright\scriptsize\tt#1}}

\def\draft{\oddsidemargin -.5truein
             \def\@oddfoot{\sl preliminary draft \hfil
             \rm\thepage\hfil\sl\today\quad\militarytime}
             \let\@evenfoot\@oddfoot \overfullrule 3pt
             \let\label=\draftlabel
             \let\marginnote=\draftmarginnote
        \def\@eqnnum{(\theequation)\rlap{\kern\marginparsep\tt\@eqnlabel}%
\global\let\@eqnlabel\@vacuum}  }

\def\numberbysection{\@addtoreset{equation}{section}
             \def\theequation{\thesection.\arabic{equation}}}

\def\underline#1{\relax\ifmmode\@@underline#1\else
             $\@@underline{\hbox{#1}}$\relax\fi}

\def\titlepage{\@restonecolfalse\if@twocolumn\@restonecoltrue\onecolumn
          \else \newpage \fi \thispagestyle{empty}\c@page\z@
             \def\thefootnote{\fnsymbol{footnote}} }

\def\endtitlepage{\if@restonecol\twocolumn \else  \fi
             \def\thefootnote{\arabic{footnote}}
             \setcounter{footnote}{0}}  
\relax

\makeatletter
\newdimen\normalarrayskip              
\newdimen\minarrayskip                 
\normalarrayskip\baselineskip
\minarrayskip\jot
\newif\ifold             \oldtrue            \def\new{\oldfalse}
\def\arraymode{\ifold\relax\else\displaystyle\fi} 
\def\eqnumphantom{\phantom{(\theequation)}}     
\def\@arrayskip{\ifold\baselineskip\z@\lineskip\z@
         \else
         \baselineskip\minarrayskip\lineskip2\minarrayskip\fi}
\def\@arrayclassz{\ifcase \@lastchclass \@acolampacol \or
\@ampacol \or \or \or \@addamp \or
       \@acolampacol \or \@firstampfalse \@acol \fi
\edef\@preamble{\@preamble
      \ifcase \@chnum
         \hfil$\relax\arraymode\@sharp$\hfil
         \or $\relax\arraymode\@sharp$\hfil
         \or \hfil$\relax\arraymode\@sharp$\fi}}
\def\@array[#1]#2{\setbox\@arstrutbox=\hbox{\vrule
         height\arraystretch \ht\strutbox
         depth\arraystretch \dp\strutbox
         width\z@}\@mkpream{#2}\edef\@preamble{\halign
\noexpand\@halignto
\bgroup \tabskip\z@ \@arstrut \@preamble \tabskip\z@ \cr}%
\let\@startpbox\@@startpbox \let\@endpbox\@@endpbox
      \if #1t\vtop \else \if#1b\vbox \else \vcenter \fi\fi
      \bgroup \let\par\relax
      \let\@sharp##\let\protect\relax
      \@arrayskip\@preamble}
\def\eqnarray{\stepcounter{equation}%
                  \let\@currentlabel=\theequation
                  \global\@eqnswtrue
                  \global\@eqcnt\z@
                  \tabskip\@centering
                  \let\\=\@eqncr
     \halign to \displaywidth\bgroup
        \eqnumphantom\@eqnsel\hskip\@centering
        $\displaystyle \tabskip\z@ {##}$%
        \global\@eqcnt\@ne \hskip 2\arraycolsep
             $\displaystyle\arraymode{##}$\hfil
        \global\@eqcnt\tw@ \hskip 2\arraycolsep
             $\displaystyle\tabskip\z@{##}$\hfil
             \tabskip\@centering
        &{##}\tabskip\z@\cr}
\begingroup\ifx\undefined\newsymbol \else\def\input#1 {\endgroup}\fi
\newfont{\hr}{msbm10}
\newfont{\ams}{msam10}

\newdimen\linethick  \linethick=0.5pt
\newdimen\hboxitspace    \hboxitspace=5pt
\newdimen\vboxitspace    \vboxitspace=5pt

\def\fr#1{%
$$\new\vcenter{\hrule height\linethick
           \hbox{\vrule width\linethick
                 \kern\hboxitspace
                 \vbox{\kern\vboxitspace
                       \hbox{$\begin{array}{c}\displaystyle#1
          \end{array}$}%
                       \kern\vboxitspace}%
                 \kern\hboxitspace
                 \vrule width\linethick}%
           \hrule height\linethick}%
$$}

\def\beq{\begin{equation}}
\def\eeq{\end{equation}}

\def\Im{{\cal I}m}
\def\Re{{\cal R}e}

\def\p{\partial}

\def\D1{{\sf D}_{1}}

\def\unitary{{\mathscr U}}
\def\hermitian{{\mathscr H}}
\def\normal{{\mathscr N}}
\def\complex{{\mathscr C}}

\def\kernel{K}
\def\lbracket{\left <}
\def\rbracket{\right >}
\def\vphcl{\varphi _{cl}}
\def\rhocl{\rho _{cl}}

\numberbysection
\hybrid

\begin{document}
\begin{titlepage}

\title{Matrix models and growth
processes: from viscous flows to the
quantum Hall effect\footnote{Based on the lectures given at
the School ``Applications of Random Matrices in Physics",
Les Houches, June 2004}}

\author{A.~Zabrodin
\thanks{Institute of Biochemical Physics,
4 Kosygina st., 119991, Moscow, Russia
and ITEP, 25 B.Cheremushkinskaya, 117259,
Moscow, Russia}}

\date{December 2004}
\maketitle

\begin{abstract}
We review the recent developments in the theory
of normal, normal self-dual and general
complex random matrices.
The distribution and correlations of the eigenvalues
at large scales are investigated in the large $N$ limit.
The $1/N$ expansion of the free energy
is also discussed.
Our basic tool is a specific Ward identity
for correlation functions (the loop equation),
which follows from
invariance of the partition function under
reparametrizations of the complex eigenvalues plane.
The method for handling the loop equation
requires the technique of boundary value problems
in two dimensions and elements of the potential theory.
As far as the physical significance of these models
is concerned, we discuss, in some detail, the recently
revealed applications to diffusion-controlled growth
processes (e.g., to the Saffman-Taylor problem) and
to the semiclassical behaviour of electronic blobs
in the quantum Hall regime.

\end{abstract}

\vfill

\end{titlepage}

\section{Introduction}

The subject matter of random matrix theory is
a matrix whose entries are
randomly distributed with some probability
density.
To put it another way, the theory deals with
statistical ensembles of matrices.
Given such an ensemble,
one is typically interested
in the distribution of eigenvalues and correlations
between them as size of the matrices, $N$, tends to infinity.
The distribution and correlation laws obtained in this way
turn out to be common to objects and systems
of very diverse nature.

The area of applications of the random matrix
theory in physics (and mathematics)
is enormously vast. It ranges from energy levels
statistics in nuclei to number theory, from quantum
chaos to string theory.
Most extensively employed
and best-understood are ensembles of hermitian
or unitary matrices, with eigenvalues being confined
either to the real line or to the unit circle.
Their applications to the level
statistics in nuclei go back to Wigner's works of
early 50-s.
For different aspects of random matrix theory,
its applications and
related topics see e.g.
\cite{Mehta}-\cite{Morozov}.

In these lectures we consider more general
classes of random matrices, with no a priori restrictions
to their eigenvalues being imposed.
The eigenvalues can be arbitrary complex numbers.
Such models are as yet less well understood but
they are equally interesting and meaningful.
As we shall see, they may exhibit
even richer mathematical structures than their
Hermitian counterparts.
Their physical applications
are also many and varied.
(A list of the relevant
physical problems and corresponding references
can be found in, e.g., \cite{list1}.)
The present lectures are based on our recent works
\cite{KKMWZ}-\cite{Agametal}
where new applications to diffusion limited growth
processes, complex analysis and quantum Hall effect
were found.

The progenitor of ensembles of matrices with general
complex eigenvalues is the statistical model of
complex matrices with the Gaussian
weight. It was introduced by Ginibre \cite{Ginibre} in 1965.
The partition function of this model is
$$
Z_N = \int [D\Phi ] \exp \left ( -\, \frac{N}{t}
\mbox{tr}\, \Phi^{\dag} \Phi \right )
$$
Here $[D \Phi ] = \prod_{ij}d({\cal R}e \, \Phi_{ij})
d({\cal I}m \, \Phi_{ij})$ is the standard volume element
in the space of $N\times N$ matrices with complex
entries $\Phi_{ij}$ and $t$ is a (real positive)
parameter.
Along with the Ginibre
ensemble and its generalizations
we also consider ensembles of normal
matrices, i.e., such that $\Phi$ commutes with
its hermitian conjugate $\Phi^{\dag}$, and normal self-dual
matrices (the definition follows below in Section 2).

Since one is primarily interested in statistics of eigenvalues,
it is natural to express the probability density
in terms of complex
eigenvalues $z_j =x_i + i y_j$
of the matrix $\Phi$.
It appears that the volume element can be represented as
$$
[D \Phi ] \propto \prod_{i<j}
\left |z_i - z_j \right |^{2\beta}\,
\prod_i d^2 z_i
$$
where $\beta =1$
for complex and normal matrices and $\beta =2$ for
normal self-dual matrices. If the statistical weight
depends on the eigenvalues only, as it is usually
assumed, the other parameters of the matrix
(often referred to as
``angular variables") are irrelevant
and can be integrated out giving an overall normalization
factor. In this case the original matrix problem
reduces to statistical mechanics of $N$ particles
with complex coordinates $z_j$ in the plane.
We thus see that even if
the matrix entries
$\Phi_{ij}$ are statistically independent, like in the
Ginibre ensemble, the eigenvalues are correlated in a
nontrivial way. Specifically, the factor
$\prod_{i<j}
\left |z_i - z_j \right |^{2\beta}$,
being equal to the exponentiated Coulomb energy
in two dimensions, means an
effective ``repelling" of eigenvalues.
This remark leads to the Dyson logarithmic
gas interpretation \cite{Dyson}, which treats the matrix
ensemble as
a two-dimensional ``plasma" of eigenvalues in an
external electric field.

At $\beta =1$,
there is another important interpretation.
Namely, the factor
$\prod_{i<j} \left (z_i - z_j \right )$
can be thought of
as coming from the Slater determinant of one-particle
fermionic states.
The averaging over matrices then turns into
averaging over the ground state of a system
of $N$ non-interacting fermions. In the case of
the Ginibre ensemble it is the system of $N$ electrons in
a uniform magnetic field at the lowest Landau level.
In the case of a spin-$\frac{1}{2}$ electron in
a non-uniform magnetic field all energy levels
split, with the only exception of the lowest one,
which remains highly degenerate.
We shall see that the normal and complex
matrix ensembles with a
non-Gaussian statistical weight are equivalent
to $N$ polarized electrons in a non-uniform magnetic
field confined to the lowest energy level.
If the degeneracy of the level equals $N$, i.e., if the
level is completely filled,
the system of $N$ electrons behaves
as an incompressible quantum Hall droplet \cite{QHdroplet}.

When $N$ becomes large some new features
emerge, which require
a different language for their adequate description,
in much the same way
as classical thermodynamics results
from statistical mechanics.
As $N\to \infty$, the eigenvalues densely fill a domain
in the complex plane with the mean
density outside it being exponentially small in $N$.
Around the edge of this domain the density steeply
drops down. The width of the transition region
tends to zero as $N\to \infty$, so that
the density profile in the direction normal
to the edge looks like a step function.
This fact allows one to introduce the
{\it support of eigenvalues}
to be the region where the mean density
of eigenvalues does not vanish as $N\to \infty$.
Typically, it is a bounded domain (or several disconnected
domains) in the complex plane.
Its shape is
determined by the probability density.

For the Ginibre ensemble,
the support of eigenvalues is the disk of radius $\sqrt{t}$
with uniform density. It is the counterpart of the
celebrated Wigner ``semicircular law".
For matrix ensembles with non-Gaussian weights
the supports are in general not circular
and not connected. Throughout these
lectures our attention is mostly restricted to the case when
the support of eigenvalues is a connected domain.
Even in this relatively simpler case,
the shape of this domain
depends on parameters of the statistical
weight in a rather complicated way.
As we shall see in Section 4, the problem to find
the support of eigenvalues from a given
statistical weight is equivalent to the inverse
problem of potential theory in two dimensions.
In most cases, solutions of the latter are not available
in an explicit form.

Nevertheless, the local dynamical law that governs
the evolution of the support of eigenvalues
under changes of parameters
of the statistical weight (like $t$ in the Ginibre
ensemble)
can be expressed in terms of the exterior Dirichlet boundary value
problem. Namely,
the edge of the support moves along gradient of a
scalar harmonic field
in its exterior, with the velocity being
proportional to the
absolute value of the gradient.
Remarkably, this growth law is known to be common to
a wide class of
diffusion-limited growth processes of which
the most popular example is viscous flow in the
Hele-Shaw cell (see \cite{RMP} for a review).
The mentioned above equivalence between the normal matrix ensemble
and the quantum Hall droplet suggests that the semiclassical
behaviour of electronic droplets in a non-uniform magnetic field
follows the same laws as the Hele-Shaw flows do.

This fact allows one to treat the model of normal or complex
random matrices as a growth problem. The advantage of this
viewpoint is two-fold. First, the hydrodynamic interpretation
makes some of the large $N$ matrix model results
more illuminating
and intuitively accessible. Second and most important,
the matrix model perspective
may help to suggest new approaches to the
long-standing growth problems. In this respect,
of special interest is the identification
of finite time singularities in some exact solutions
to the Hele-Shaw flows with critical points of the normal
and complex matrix models.

At last, a few words about the organization of the lectures.
The material that follows
can be divided into three parts.
The first one (Section 2) can be regarded as a continuation of the
introduction. We define the main matrix ensembles to be considered
and give their physical interpretations.
The second part (Section 3) contains exact results
valid at any finite $N$. We outline
the integrable structure of the
normal and complex models at $\beta =1$ (the Hirota
relations for the partition function,
the orthogonal polynomials technique and the Lax representation).
In addition, we derive the exact relation between
correlation functions of the eigenvalue densities
(referred to as the loop equation) which holds for arbitrary
values of $\beta$. In the third part (Sections 4 and 5)
we examine the large $N$ limit of the models of random
matrices with complex eigenvalues and discuss the applications
to the growth processes and to the semiclassical electronic
droplets in magnetic field. The Appendices contain
technical details of some proofs and calculations.

\section{Some ensembles of random matrices
with complex eigenvalues}

We consider square random matrices $\Phi$
of size $N$
with complex entries $\Phi_{ij}$
subject to certain constraints depending on
the particular ensemble.
Some ensembles of random matrices are listed in the following
table:

\begin{center}

\begin{tabular}{|c|c|c|c|}
\hline
Ensemble & Notation & Condition & Dimension  \\
\hline
&&&\\
Hermitian & $\hermitian$ & $\Phi^{\dag} =\Phi$ & $N^2$   \\
&&&\\
Unitary   & $\unitary$   & $\Phi^{\dag}\Phi =1$ &$N^2$    \\
&&&\\
\hline
&&&\\
Normal   & $\normal$ & $[\Phi^{\dag}, \Phi ]=0$ &
$N^2 \!+ \! N$  \\
&&&\\
$\begin{array}{l}
\mbox{{\small Normal}}\\
\mbox{{\small self-dual}}
\end{array}$
& $\normal ^0$ &
$\begin{array}{c} [\Phi^{\dag}, \Phi ]=0\\
\Phi \; \mbox{{\small self-dual}}
\end{array}$ & $\frac{1}{2}N^2 +N$ \\
&&&\\
Complex & $\complex$ & {\small none} & $2N^2$  \\
&&&\\
\hline
\end{tabular}

\end{center}

\noindent
The first two matrix ensembles,
$\hermitian$ and $\unitary$, are the most popular ones.
They are given here just for comparison.
Eigenvalues of matrices from $\hermitian$ and $\unitary$
are confined to the real axis and to the unit circle respectively.
The last three ensembles (which are the main subject of
these lectures) do not imply
any a priori restrictions on eigenvalues of the matrices.
Normal matrices are defined by the constraint that they
commute with their adjoint.
The ensemble $\normal^0$ is defined for $N$ even only.
The meaning of the condition ``$\Phi$ is self-dual" is explained
below in this section. By dimension of the ensemble we mean
the {\it real} dimension of the matrix variety.

Throughout this paper we consider the probability densities
of the form $P(\Phi )\propto e^{{\rm tr} W(\Phi )}$, where
the function $W(\Phi )$ (often called the potential of
the matrix model) is a matrix-valued
function of $\Phi$ and $\Phi^{\dag}$ such that
$(W(\Phi ))^{\dag} =W(\Phi )$. This form is similar
to the one usually employed in Hermitian and unitary ensembles.
The partition function is defined as the integral over matrices
from one or another ensemble:
\beq\label{partition}
Z_N = \int [D \Phi ] e^{\mbox{tr}\, W(\Phi )}
\eeq
Summing over $N$ with a suitable weight, one may also
define the grand canonical ensembles
corresponding to (\ref{partition}) but we do not
pursue this possibility here.

We need to specify
the integration measure $[D\Phi ]$ and the potential
$W(\Phi )$.
Given the measure and the potential, one is usually interested
in the distribution and correlations of the eigenvalues.
In general, they can be distributed
on the real line for $\hermitian$, on the unit circle
for $\unitary$ and on the whole complex plane
for $\normal$, $\normal^0$ and $\complex$.

\subsection{Integration measures}

The integration measure has the most simple form
for the ensemble of general complex matrices:
$$
[D \Phi ]=\prod_{i,j =1}^{N}
d(\Re \, \Phi_{ij})\, d (\Im \, \Phi_{ij})
$$
This measure is additively invariant and multiplicatively
covariant, i.e.
for any fixed (nondegenerate) matrix $A\in \complex$ we have
the properties
$[D(\Phi +A)] =[D\Phi ]$ and
$[D(\Phi A)] =[D(A\Phi )]=|\det A|^{2N} [D\Phi ]$.
The first one is obvious, to prove the second one is an easy
exercise. It is clear that the measure is invariant under
transformations of the form $\Phi \to U^{\dag}\Phi U$ with a unitary
matrix $U$ (``rotations" in the matrix space).

The measure for $\normal$ is induced by the
standard flat metric in $\complex$,
$$
||\delta \Phi ||^2 =\mbox{tr} \,
(\delta \Phi \delta \Phi^{\dag}) =\sum_{ij} |\delta \Phi_{ij}|^2
$$
via the embedding
$\normal \subset \complex$.
Here $\normal$ is regarded as a hypersurface in $\complex$ defined by
the quadratic relations $\Phi \Phi^{\dag}=\Phi^{\dag}\Phi$.
The measure for the ensemble $\normal ^0$ is defined in a similar
way.

As usual in matrix models, we would like to integrate out
the ``angular'' variables and to express the integration measure
through eigenvalues of the matrices.

\paragraph{The measure for $\normal$ through
eigenvalues \cite{Mehta,normal}.}
We derive the explicit representation of the measure in terms
of eigenvalues in three steps:

\begin{itemize}
\item[1.]
Introduce coordinates in $\normal \subset \complex$.
\item[2.]
Compute the inherited metric on $\normal$ in these
coordinates: $||\delta \Phi ||^2 =g_{\alpha \beta}
d\xi^{\alpha}d\xi^{\beta}$.
\item[3.]
Compute the volume element
$[D\Phi ]=\sqrt{|\det g_{\alpha \beta}|}
\prod_{\alpha} d\xi^{\alpha}$.
\end{itemize}

\noindent
{\it Step 1: Coordinates in $\normal$}.
For any matrix $\Phi$, the matrices
$H_1 =\frac{1}{2}(\Phi +\Phi^{\dag})$,
$H_2 =\frac{1}{2i}(\Phi -\Phi^{\dag})$
are Hermitian.
The condition
$[\Phi , \Phi^{\dag}]=0$ is equivalent to
$[H_1 , H_2]=0$.
Thus $H_{1,2}$ can be simultaneously diagonalized
by a unitary matrix $U$:
$$
\begin{array}{l}
H_1 =UXU^{\dag}\,,
\quad X=\mbox{diag}\, \{ x_1 , \ldots , x_N\}\\
H_2 =UYU^{\dag}\,,
\quad \, \, Y=\mbox{diag}\, \{ y_1 , \ldots , y_N\}
\end{array}
$$
Introduce the diagonal matrices
$Z=X+iY$, $\bar Z=X-iY$ with diagonal elements
$z_j =x_j +iy_j$ and $\bar z_j =x_j - iy_j$ respectively.
Note that $z_j$ are eigenvalues of $\Phi$.
Therefore, any $\Phi \in \normal$
can be represented as
$$
\Phi =UZU^{\dag}
$$
where $U$ is a unitary matrix and $Z$ is the diagonal
matrix with eigenvalues of $\Phi$ on the diagonal.
In fact normal matrices can be equivalently defined
by the property of being the most general matrices
that can be diagonalized by a unitary transformation.
The matrix $U$ is defined up to multiplication by a diagonal
unitary matrix from the right:
$U \rightarrow U \, U_{{\rm diag}}$.
The dimension of $\normal$ is thus
$$
\mbox{dim}\, (\normal )=
\mbox{dim}\, (\unitary ) -
\mbox{dim}\, (\unitary_{{\rm diag}} )+
\mbox{dim}\, (\complex_{{\rm diag}} )=
N^2 -N + 2N \, =\, N^2 +N
$$

Let us make a remark that the naive counting of
the number of constraints in the condition
$[\Phi , \Phi^{\dag}]=0$ leads to a wrong result
for the dimension of $\normal$.
On the first glance,
this condition gives $N^2 -1$ independent constraints. Indeed,
set $H=[\Phi , \Phi^{\dag}]$. Then the conditions
$H_{lk}=0$ for $l<k$ give $N(N\! -\! 1)$ real constraints
and the conditions $H_{kk}=0$ give $N-1$ real constraints
(because $\mbox{tr}\, H =0$ identically), in total
$N^2 -1$ constraints.
We thus observe that
$\mbox{dim}\, (\normal ) \neq \mbox{dim} \, (\complex )
-(N^2 \! -\! 1)$.
Therefore, there are only $N^2 -N$ independent constraints
among the $N^2 -1$ equations
$[\Phi , \Phi^{\dag}]=0$. This fact can be easily illustrated
by the example of $2\times 2$ matrices.

\noindent
{\it Step 2: The induced metric}.
Since $\Phi =UZU^{\dag}$, the variation is
$\delta \Phi =U(\delta u \cdot Z + \delta Z + Z \cdot \delta
u^{\dag})U^{\dag}$,
where
$\delta u^{\dag}=U\delta U^{\dag}=-\delta u^{\dag}$.
Therefore,
$$
||\delta \Phi ||^2 =\, \mbox{tr}\,
(\delta \Phi \delta \Phi^{\dag})
=\, \mbox{tr}\, (\delta Z \delta \bar Z) +2\, \mbox{tr}\,
(\delta u Z \delta u \bar Z \! -\! (\delta u)^2 Z \bar Z)
$$
$$
=\sum_{j=1}^{N}|\delta z_j |^2 + 2 \sum_{j<k}^{N}
|z_j -z_k |^2 \, |\delta u_{jk}|^2
$$
(Note that $\delta u_{jj}$ do not enter.)
This is the square of the  line element
$||\delta \Phi ||^2 =
g_{\alpha \beta}\delta \xi^{\alpha}\delta\xi^{\beta}$.

\noindent
{\it Step 3. The volume element}.
We see that the metric $g_{\alpha \beta}$ is diagonal
in the coordinates
$\Re (\delta z_j)$,
$\Im (\delta z_j)$,
$\Re (\delta u_{jk})$,
$\Im (\delta u_{jk})$ with $1\leq j<k\leq N$, so the determinant
of the diagonal matrix $g_{\alpha \beta}$ is
easily calculated to be
$|\det g_{\alpha \beta}|=2^{N^2 -N}
\prod_{j<k}^{N} |z_i -z_k |^4$. Therefore,
\beq\label{measure1}
[D \Phi ]\propto [D U]' \, |\Delta_N (z_1 , \ldots , z_N )|^2
\prod_{j=1}^{N} d^2 z_j
\eeq
where $d^2 z\equiv dx dy$ is the
flat measure in the complex plane,
$[DU]' = [DU]/ [DU_{{\rm diag}}]$
is the invariant measure on $\unitary / \unitary _{{\rm diag}}$,
and
\beq\label{Vandermonde}
\Delta_N (z_1 , \ldots , z_N)=\prod_{j>k}^{N} (z_j \! -\! z_k )
=\det_{N\times N} (z_{j}^{k-1})
\eeq
is the Vandermonde determinant.

Similarly to the ensemble
$\hermitian$ of Hermitian matrices, the measure
(\ref{measure1}) contains the squared modulus of the
Vandermonde determinant. The difference is that the eigenvalues
are complex numbers. The statistical model of normal random
matrices was studied in \cite{normal,Oas}.

\paragraph{Normal self-dual matrices.}
Let $\Gamma$ be the matrix
$$
\Gamma =\left ( \begin{array}{cccccc}
0& 1 &&&&\\
-1 & 0 &&&&\\
&& 0&1 &&\\
&& -1&0 &&\\
&&&& \ldots &\\
&&&&& \ldots
\end{array}
\right ), \;\;\;
\Gamma^2 =-1
$$
(all other entries are zero).
A complex matrix $\Phi$ is called {\it self-dual} if
$\Gamma \Phi^{{\rm T}} \Gamma =-\Phi$ (the superscript
${\rm T}$ means transposition).
The size of a self-dual matrix
is thus an even number.
It can be shown that eigenvalues of
self-dual matrices always come in pairs:
the diagonal form of $\Phi$ is
$$
Z=\mbox{diag}\, \{ z_1 , z_1 , \, z_2 , z_2 , \ldots ,
z_N , z_N \}
$$
As is easy to verify, the condition that $\Phi$ is
self-dual is equivalent to the condition that
the matrix $\Gamma \Phi$ is anti-symmetric.

{\it Normal self-dual} matrices are
parameterized as
$\Phi =UZ U^{\dag}$
with $Z$ as above, where $U$ is unitary and symplectic:
$U^{\dag}U=1$, $U^{{\rm T}} \Gamma U =\Gamma$. In other words,
$U$ belongs to the maximal compact subgroup in the
complex group of symplectic matrices $\mbox{Sp}(N)$.
The latter is known to have real dimension $4N^2 + 2N$,
with the dimension of the maximal compact subgroup being twice
less. Therefore, similarly to the calculation for
normal matrices, $\mbox{dim}(\normal^0 ) =
2N^2 +N -N +2N = 2N^2 + 2N$ (the value given in the table
above corresponds to $N$ replaced by $N/2$).
The integration measure appears to be
\beq\label{measure2}
[D \Phi ]\propto [D U]' \, |\Delta_N (z_1 , \ldots , z_N )|^4
\prod_{j=1}^{N} d^2 z_j
\eeq
(for $2N \times 2N$ matrices). Note that the module of the Vandermonde
determinant enters in the fourth degree.
The statistical model of normal self-dual matrices was
discussed in \cite{Hastings}.

\paragraph{The measure for $\complex$ through eigenvalues.}
A complex matrix $\Phi$ with eigenvalues $z_1 , \ldots , z_N$
can be decomposed as
$$
\Phi =U(Z+R)U^{\dag}
$$
where $Z=\mbox{diag}\, \{ z_1 , \ldots , z_N\}$ is
diagonal, $U$ is unitary, and $R$ is strictly upper triangular, i.e.,
$R_{ij}=0$ if $i\geq j$. These matrices are defined up to a
``gauge transformation":
$U\to U\, U_{{\rm diag}}$,
$R\to U^{\dag}_{{\rm diag}} \, R \, U_{{\rm diag}}$.
It is not so easy to see that the measure factorizes.
This requires some work, of which the key
step is a specific ordering of the independent variables.
The final result is:
\beq\label{measure3}
[D \Phi ] \propto
[DU]' \, \left ( \prod_{k<l} d^2 R_{kl}\right )
|\Delta_{N}(z_i)|^2 \prod_{j=1}^{N} d^2 z_j
\eeq
The details can be found in the Mehta book \cite{Mehta}.

\subsection{Potentials}

For the ensembles $\normal$, $\normal ^0$ the
``angular variables" (parameters of the unitary matrix $U$)
always decouple after taking the trace
$\mbox{tr}\, W(\Phi )=\sum_j W(z_j)$,
so the potential $W$ can be a function
of $\Phi$, $\Phi^{\dag}$ of a general form
$W (\Phi ) =\sum a_{nm} \Phi^n (\Phi^{\dag})^m$.
Two important particular cases arise if the potential is:
\begin{itemize}
\item
Axially symmetric, $W(\Phi )=W_0 (\Phi \Phi^{\dag})$.
In this case the $N$-fold integral essentially
reduces to ordinary ones, and so some basic results
become available in a quite explicit form.
\item
Harmonic on the background of $\Phi \Phi^{\dag}$, i.e.,
$W(\Phi )=-\Phi \Phi^{\dag} +V(\Phi ) +
\bar V(\Phi ^{\dag})$.
In what follows, we call it {\it quasiharmonic}.
Here $V(z)$ is an {\it analytic} function of $z$ in some
domain containing the origin and
$\bar V(z)=\overline{V(\bar z)}$.
In terms of the eigenvalues, the quasiharmonic potential is
\beq\label{quasiharm}
W(z )=-|z|^2 +V(z ) +
\overline{V(z)}
\eeq
This case is particularly important for applications.
The normal matrix model with quasiharmonic potentials
bears some formal similarities with the model of two
coupled Hermitian matrices \cite{DKK} and the
matrix quantum mechanics in the singlet sector \cite{AKK}.
\end{itemize}
The partition function reduces to
\beq\label{Z}
Z_N =\int |\Delta_N (z_i)|^{2\beta}
\prod _{j=1}^{N}
e^{W(z_j)} d^2 z_j
\eeq
where $\beta =1$ for $\normal$ and $\beta = 2$ for $\normal^0$.
From now on this formula is taken as the definition of the
partition function. Comparing to (\ref{partition}), we
redefine $W \to W/\beta$ and ignore a possible
$N$-dependent normalization factor.
One may also
consider this integral for arbitrary values of $\beta$.

The choice of the potential for the
ensemble $\complex$ is more restricted.
For a general potential,
the matrix $U$ in
$\Phi =U(Z+R)U^{\dag}$
still decouples but $R$ does not. The problem becomes too
complicated.
An important particular case, when $R$ nevertheless
decouples is the quasiharmonic potential
(\ref{quasiharm}).
Indeed,
$
\mbox{tr}\, (\Phi \Phi^{\dag})=
\mbox{tr}\, (Z\bar Z)+
\mbox{tr}\, (R R^{\dag})$,
$\mbox{tr}\, (\Phi ^n)=
\mbox{tr}\, (Z+R)^n =\mbox{tr}\, Z^n$,
and so
\beq\label{measure4}
\int_{\complex} [D\Phi ] e^{\mbox{tr}\, W(\Phi )}=
C_N
\int |\Delta (z_i)|^2 \prod_k e^{W(z_k)}d^2 z_k
\eeq
where $C_N$ is an $N$-dependent normalization factor
proportional to the gaussian integral
$\int [D R] e^{-\mbox{tr}\, (RR^{\dag})}$.

As an example, let us consider the quadratic potential:
$$
W(z)=-\sigma |z|^2 +2\Re \, (t_1 z +t_2 z^2 )\,,
\quad \sigma >0
$$
The ensemble $\complex$ ($\beta =1$) with this potential
is known as {\it the Ginibre-Girko ensemble} \cite{Ginibre,Girko}.
In this case the partition function (\ref{Z})
can be calculated exactly \cite{FGIL}:
$$
Z_N =
 Z_{N}^{(0)}
(\sigma^2 -4|t_2|^2 )^{-N^2 /2}
\exp \left (N\,
\frac{t_{1}^{2}\bar t_2
+\bar t_{1}^{2}t_2 +\sigma |t_1|^2}{\sigma^2 -4|t_2|^2}
\right )
$$
where
$$
Z_{N}^{(0)}
=\sigma^{(N^2 -N)/2} \pi^N \prod_{k=1}^{N}k!
$$
To the best of our knowledge, there are no exact results for
2D integrals of this type  with $|\Delta (z_i)|^{2\beta}$
for other values of $\beta$,
even for the
pure Gaussian weight.

Coming back to the general case,
we note that some integrals considered above may diverge.
In the most important case of quasiharmonic potential,
the integral
$$
\int e^{-|z|^2 +V(z)+\overline{V(z)}}d^2 z
$$
converges for potentials
$
V(z)=\alpha z^2 +\beta z +\sum_i \mu_i \log
(z-a_i )
$
with $|\alpha |<\frac{1}{2}$ and $\mu_j >-1$
but it always diverges if $V(z)$ is a polynomial of degree
$\geq 3$.
As usual in matrix models, really interesting
science begins when integrals diverge!
Let us say a few words about
how one should understand divergent integrals.
The conventional viewpoint
is to treat all cubic and higher degree terms in the potential
as a small perturbation.
The integral is then regarded
as a perturbative series for a theory which is
believed to be well-defined
on the nonperturbative level.
The nonperturbative definition can be achieved
either by introducing an ad hoc cutoff or via more sophisticated
methods in the spirit of the Marinari-Parisi
approach \cite{MP} (the ``stochastic stabilization'').
As far as the large $N$ limit is concerned,
the integral for the partition function can be
{\it defined} through the expansion around the saddle
point. This gives the $1/N$-expansion
$$
\log Z_{N}\sim \sum_h N^{-h}F^{(h)}
\quad \quad \mbox{as $N\to \infty$}
$$
Even if the integral for
$Z_N$ diverges, each term of the $1/N$ expansion
is often well-defined.

\subsection{Physical interpretations}

The ensembles of random matrices
appear to be mathematically equivalent to
some important model systems of statistical and quantum mechanics.
They are:
\begin{itemize}
\item
The 2D Coulomb plasma (any $\beta$)
\item
Non-interacting fermions ($\beta =1$)
\item
Electrons in magnetic field ($\beta =1$)
\end{itemize}
The equivalence holds for any finite $N$.
The first two interpretations are standard and well known.
The third one can be regarded as a specification
of the second one for models with eigenvalues
distributed over the whole complex plane.
Remarkably, in this very case the noninteracting fermions
picture (which is rather formal for $\hermitian$ and $\unitary$)
acquires a very interesting physical content related to the
quantum Hall effect.

\paragraph{The Dyson gas picture.}
This interpretation, first suggested by Dyson \cite{Dyson}
for the unitary, symplectic and orthogonal matrix ensembles,
relies on
rewriting $|\Delta_N (z_i)|^{2\beta}$ as
$\exp \Bigl (\beta \sum_{i\neq j}\log |z_i \! - \! z_j | \Bigr )$.
Clearly, the integral (\ref{Z}) looks then
exactly as the partition function of the
2D Coulomb plasma (often called the Dyson gas) at
``temperature" $1/\beta$, in the external
electric field:
\beq\label{ZE}
Z_N =\int e^{-\beta E(z_1 , \ldots , z_N)}
\prod d^2 z_j
\eeq
The eigenvalues play the role of the 2D Coulomb charges.
The energy is
\beq\label{energy}
E=-\sum_{i<j}\log |z_i -z_j |^2
-\beta^{-1} \sum_j W(z_j)
\eeq
The first sum is the Coulomb interaction energy, the
second one is the energy due to the external field.
For the ensembles $\hermitian$ and $\unitary$
the charges are confined to dimension 1
(the real line or the unit circle)
but interact as 2D Coulomb charges. So, the Dyson gas picture
for the ensembles $\normal$, $\normal^0$ and $\complex$
looks even more natural. The Dyson gas interpretation becomes
especially helpful in the large $N$ limit, where it allows
one to apply thermodynamical arguments.

\paragraph{Non-interacting fermions ($\beta =1$).}
Given any system of polynomials of the form
$P_n(z)=z^n + \mbox{lower degrees}$,
the Vandermonde determinant can be written as
$\Delta_N(z_i)=\det (z_{j}^{k-1})=\det
(P_{k-1}(z_j))$.
Let us rewrite the statistical weight
as
$$
|\Delta_{N}(z_i)|^2 \prod_{j=1}^{N}e^{W(z_j)}
=\left | \Psi (z_1 , \ldots , z_N)\right |^2
$$
where $\Psi (z_1 , \ldots , z_N) =\det _{N\times N}
\left ( P_{k-1}(z_j)e^{W(z_j)}\right )$
is the (unnormalized) wave function of $N$
non-interacting fermions (the Slater determinant),
with one-particle wave functions being
$\psi_k (z) =P_{k-1}(z)e^{W(z)}$.
The partition function is the normalization integral:
$$
Z_N =\int |\Psi |^2 \prod_i d^2 z_i
$$
For the ensembles $\normal$ and
$\complex$ (with quasiharmonic potential) the wave function
$\Psi$ has a direct
physical meaning as a wave function of
2D electrons in
magnetic field.

\paragraph{Electrons in the plane in magnetic field.}
Consider a charged particle with spin $\frac{1}{2}$
(electron) moving in the plane in a strong
(not necessarily uniform)
magnetic field $B = B(x,y)$ orthogonal to the plane.
The Pauli Hamiltonian reads
$$
\hat H=\frac{1}{2m}\left ( ( i\hbar \nabla
+\vec A)^2 - \hbar \sigma_3 B \right )
$$
Here $m$ is mass of the particle,
$\sigma_3 =\mbox{diag}\, (1, -1)$
is the Pauli matrix, $B=\p_x A_y -\p_y A_x$.
In 2D, the complex notation is convenient:
$A=A_x -iA_y$, $\bar A=A_x +iA_y$,
so that
$B=i(\bar \p A - \p \bar A)$.

For a uniform magnetic field, $B=B_0$,
one can choose the gauge
$A=\frac{B_0}{2i}\, \bar z$.
Solving the Schrodinger equation
$\hat H \psi =E\psi$,
one gets Landau levels:
$$
E_n =\frac{\hbar B_0}{m}\left ( n+\frac{1}{2} - s \right )\,,
\;\;\;\;\;
n =0, 1, 2 , \ldots \; , \;\; s=\pm \frac{1}{2}
$$
The gap between the levels
is proportional to $B_0/m$. Each level is highly degenerate.
The wave functions at the level $E=0$ are
$$
\psi_n =z^n\exp \left ( -\frac{B_0}{4\hbar}|z|^2\right )
$$
We note in passing that the system of fermions in the magnetic
field at the lowest energy level
admits a collective field theory description which was
discussed in the literature in different contexts
(see e.g. \cite{QHdroplet,Hastings2,BCR}).

\begin{figure}[tb]
\epsfysize=5.5cm
\centerline{\epsfbox{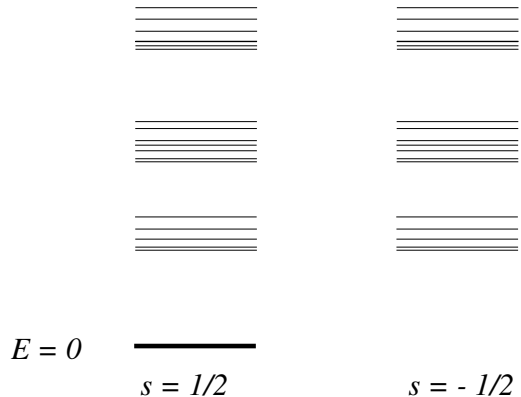}}
\caption{\sl The energy levels of a spin-$\frac{1}{2}$
electron in a non-uniform magnetic field (schematically).
The lowest
level $E=0$ remains highly degenerate. }
\label{fi:levels}
\end{figure}

Let us turn to the problem with a nonuniform magnetic field.
Choose the gauge
$A=i\, \p W$ with a real-valued function $W$,
then
$B=-2\,  \p \bar \p W$
and $\mbox{div}\, \vec A =\bar \p A+\p \bar A =0$.
Therefore,
$$
(i\hbar \nabla +\vec A)^2 =
-4\hbar^2 \p \bar \p +
2i\hbar ((\bar \p W) \p + (\p W)\bar \p )+ |\p W|^2
$$
$$
=(2\hbar \p +\p W)
(-2\hbar \bar \p +\bar \p W) -2\hbar \, \p \bar \p W
$$
The Hamiltonian can be represented as the 2$\times$2 matrix
$$
\hat H=\left (
\begin{array}{cc}
H_+ & 0\\ 0 & H_- \end{array}\right )
$$
where
$2mH_{\pm}=(i\hbar \nabla +\vec A)^2
\pm 2\hbar \, \p \bar \p W$.
In general, the spectral problem
for this Hamiltonian does not admit an explicit solution.
However, the level $E=0$ is very special.
Note that $H_{+}$ factorizes:
$H_+ =(2\hbar \p +\p W)
(-2\hbar \bar \p +\bar \p W)$,
so exact wave functions at the level $E=0$
can be found by solving the {\it first order} equation
$$
H_{+}\psi = (2\hbar \bar \p -\bar \p W)\psi =0
$$
The general solution is
\beq\label{psi}
\psi (z)= P(z)\exp \left (\frac{1}{2\hbar}W(z)\right )
\eeq
where $P(z)$ is an arbitrary
{\it holomorphic} polynomial.
The zero energy level remains to be highly degenerate
even in the nonuniform magnetic field
(Fig.~\ref{fi:levels}). This fact was first
observed in \cite{Ahar-Casher}.

To find degeneracy of the level, we solve the Poisson equation
$\Delta W=-2B$,
\beq\label{mag1}
W(z)=-\frac{1}{\pi}\int \log |z-\zeta |B(\zeta )d^2 \zeta
\eeq
and observe that
$W(z)$ tends to
$-\frac{\phi}{\pi}\, \log |z|$ as $|z|\to \infty$,
where
$
\phi =\int B\, d^2 z
$
is the total magnetic flux.
If $P(z)$ is of degree $n$, the asymptotics of $\psi$
for $|z|$ large is
$$
\psi =P(z)\, e^{\frac{1}{2\hbar}W(z)} \rightarrow
z^n \, |z|^{-\phi / \phi_0}
$$
where $\phi_0 =2\pi \hbar$ is the flux quantum.
We require the wave functions to be normalizable, i.e,
$\int |\psi |^2 d^2 z < \infty$ that means
$n< \phi /\phi_0 -1$.
Therefore,
$$
n_{{\rm max}} =\left [ \phi /\phi_0 \right ] -1
$$
($[\ldots ]$ is the integer part), and the degeneracy
is equal to the number of flux quanta in the total flux:
$$
N=\left [ \phi / \phi_0 \right ]
$$
If the Coulomb forces can be ignored,
the wave function of $N$ electrons in the plane in
the magnetic field at the lowest energy level is constructed
as the $N\times N$
Slater determinant of the functions of the type
(\ref{psi}) with polynomials of different degrees.

The situation when the lowest energy level $E=0$ is
completely filled, i.e.,
$N=n_{{\rm max}}$, is the (integer) quantum Hall (QH) regime.
The notion of the QH droplet \cite{QHdroplet}
implies that the electronic liquid is incompressible, i.e.,
all states at the lowest energy level are occupied.
We come to the following conclusion:
the QH droplet consisting of $N$ electrons
(in general, in a non-uniform magnetic field)
is equivalent to the ensemble of normal $N\times N$ matrices.

\section{Exact results at finite $N$}

\subsection{Correlation functions: general relations}

The main objects to be determined in random matrix
models are correlation functions. In general, they are mean values
of scalar-valued functions of matrices.
The mean value of such a function
$A(\Phi )$ of the matrix $\Phi$ is defined,
with the help of the statistical weight, in the usual way:
$$
\lbracket A\rbracket =
\frac{\displaystyle{\int [D\Phi ]A(\Phi )
e^{\mbox{tr}\, W(\Phi )}}}{\displaystyle{\int [D\Phi ]
e^{\mbox{tr}\, W(\Phi )}}}
$$
We shall consider functions that depend on
eigenvalues only -- for example, traces
of matrices.
Correspondingly, typical correlators which we are
going to study are mean values of products of traces:
$\lbracket \mbox{tr}\, f(\Phi )\rbracket$,
$\lbracket \mbox{tr}\, f_1(\Phi )
\, \mbox{tr}\, f_2(\Phi ) \rbracket $
and so on.
Clearly, they
are represented as integrals over eigenvalues. For example,
$$
\lbracket \mbox{tr}\, f(\Phi )\rbracket =
\frac{N \displaystyle{\int |\Delta_N (z_i)|^{2\beta}}
f(z_1) \prod_{j=1}^{N} e^{W(z_j)}d^2 z_j}{\displaystyle{
\int |\Delta_N (z_i)|^{2\beta}}
\prod_{j=1}^{N} e^{W(z_j)}d^2 z_j}
$$
Here, $f(\Phi )=f(\Phi, \Phi^{\dag})$
is any function of $\Phi$, $\Phi^{\dag}$ which is regarded as
the function $f(z_i)=f(z_i, \bar z_i)$
of the complex argument $z_i$ (and $\bar z_i$) in the r.h.s.
A particularly important example is the density function
defined as
\beq\label{density1}
\rho (z)=\sum_j \delta (z-z_j)=\mbox{tr}\, \delta (z-\Phi )
\eeq
where $\delta (z)$ is the two dimensional $\delta$-function.
As it immediately follows from the
definition, any correlator of traces is expressed through
correlators of $\rho$:
\beq\label{trf}
\lbracket \mbox{tr}\, f_1 (\Phi ) \, \ldots \,
\mbox{tr}\, f_n (\Phi )\rbracket
=\int \lbracket \rho (z_1) \ldots \rho (z_n)\rbracket
f_1 (z_1 ) \ldots f_n (z_n ) \prod_{j=1}^{n}d^2 z_j
\eeq
Instead of correlations of density it is often convenient
to consider correlations of the field
\beq\label{potential1}
\varphi (z)=-\beta \sum_j \log |z-z_j |^2
=- \beta \log \left | \det (z-\Phi ) \right |^2
\eeq
from which the correlations of density can be found
by means of the relation
\beq\label{rhophi}
4\pi \beta \rho (z)=-\Delta \varphi (z)
\eeq
Clearly, $\varphi$ is the 2D Coulomb potential
created by the eigenvalues (charges).
As it directly follows from the definitions,
$$
\lbracket \rho (z) \rbracket _{N} =N\,
\frac{Z_{N-1}}{Z_N} \lbracket e^{W(z)-\varphi (z)}
\rbracket _{N-1}
$$
where $\lbracket \ldots \rbracket_N$ means the
expectation value
in the ensemble of $N\times N$ matrices.

Handling with multipoint correlation functions,
it is customary to pass to their {\it connected parts}.
For example, in the case of 2-point functions, the
connected correlation function is defined as
$$
\lbracket \rho (z_1 )\rho (z_2)\rbracket _{c}\equiv
\lbracket \rho (z_1 )\rho (z_2)\rbracket -
\lbracket \rho (z_1) \rbracket
\lbracket \rho (z_2) \rbracket
$$
The connected multi-trace correlators are expressed
through the connected density correlators by the same
formula (\ref{trf}) with $\lbracket \rho (z_1)\ldots
\rho (z_n) \rbracket_c$ in the r.h.s.
The connected part of the
$(n+1)$-point density correlation function is given by the
linear response of the $n$-point one to a small variation
of the potential.
More precisely,
the following variational formulas
hold true:
\beq\label{var}
\lbracket \rho (z) \rbracket =
\frac{\delta \log Z_N}{\delta W(z)}\,,
\;\;\;\;
\lbracket \rho (z_1 )\rho (z_2)\rbracket _{c}=
\frac{\delta \lbracket \rho (z_1)\rbracket }{\delta W(z_2)}=
\frac{\delta^2 \log Z_N}{\delta W(z_1) \delta W(z_2)}
\eeq
Connected multi-point correlators are higher
variational derivatives of $\log Z_N$.
These formulas follow from the fact that
variation of the partition function over a general
potential $W$ inserts $\sum_i \delta (z-z_i)$ into
the integral.
Let us stress that these formulas
are exact for any finite $N$.

For a later use, we mention the formula
\beq\label{expf}
\lbracket e^{\mbox{tr} \, f (\Phi )}\rbracket
=\exp \left ( \sum_{k=1}^{\infty} \frac{1}{k!}
\lbracket (\mbox{tr}\, f(\Phi ))^k \rbracket_c \right )
\eeq
which immediately follows from the expansion
$$
\log
\lbracket e^{\mbox{tr} \, f }\rbracket
=\log Z_N (W+f)- \log Z_N (W)
$$
$$
=\, \int \frac{\delta \log Z_N}{\delta W(\zeta )}
\, f(\zeta ) d^2 \zeta
\, + \, \frac{1}{2!}
\int \frac{\delta^2 \log Z_N}{\delta W(\zeta )\delta W(\zeta ' )}
\, f(\zeta )f(\zeta ') d^2 \zeta d^2 \zeta ' \,
+\ldots
$$

\subsection{Integrable structure of the
$\normal$ and $\complex$ ensembles ($\beta =1$)}

The partition function (\ref{Z}) for $\beta =1$,
$$
Z_N =\int |\Delta_N (z_i)|^2 \prod_{j=1}^{N}
e^{W(z_j)}d^2 z_j
$$
regarded as a function of $N$ and Taylor
coefficients of the potential $W$,
has remarkable properties, which we briefly review
below.

\paragraph{Determinant representation.}
The following simple but important determinant
representation holds true:
\beq\label{det}
\phantom{\int} Z_N =N! \, \det_{N\times N} (C_{ij})\,,
\quad  1\leq i,j\leq N
\eeq
where
\beq\label{det1}
C_{ij}=\int z^{i-1}\bar z^{j-1} e^{W(z)}d^2z
\eeq
is the ``matrix of moments".
The proof is almost a repetition of the
corresponding proof for the hermitian model.
Complexity of eigenvalues does not cause any
difficulties. For completeness, the detailed
proof is given in the Appendix.

\paragraph{Hirota relations.}
Let us apply the determinant formula to
$\lbracket \det (\lambda -\Phi )\rbracket$:
$$
\lbracket \det (\lambda -\Phi )\rbracket
=\frac{1}{Z_N}\int |\Delta_N|^2 \prod_j (\lambda -z_j)
e^{W(z_j)}d^2 z_j
=\, \frac{1}{Z_N}\det \left [
z^{i-1}\bar z^{j-1} (\lambda \!-\! z) e^{W} d^2 z \right ]
$$
Comparing this with the determinant representation of
$Z_N$, we see that
$\lbracket \det (\lambda -\Phi )\rbracket =
\lambda^N Z_{N}^{-1}
\det \left [ C_{ij}\! - \! \lambda^{-1}
C_{i+1, j}\right ]$. Taking appropriate
linear combinations of the lines, one can reduce
this determinant to the determinant of a matrix
which differs from $C_{ij}$ only in the last line.
Similarly, in the determinant
representation of
$\lbracket \det (\lambda_1 -\Phi )
\det (\lambda_2 -\Phi )\rbracket$
only two last lines change.
Now, some standard
identities for determinants lead
to the following relations:
\beq\label{prehir1}
(\lambda_1 \! -\! \lambda_2) \left < \det (\lambda_1 \! -\! \Phi )
\det (\lambda_2 \! -\! \Phi )\right >
\left < \det (\lambda_3 \! -\! \Phi )\right >
+\, \mbox{cyclic perm-s of (123)} \, = \, 0
\eeq
\beq\label{prehir2}
\left < \left | \det (\lambda \! -\! \Phi )\right |^2 \right >_N
-\left |\left < \det (\lambda \! -\! \Phi )\right >_N \right |^2
=\, \frac{N}{N+1}\, \frac{Z_{N+1}Z_{N-1}}{Z_N^2}
\left <
\left | \det (\lambda \! -\! \Phi )\right |^2 \right >_{N-1}
\eeq
Write
$$
\left < \det (\lambda \! -\! \Phi ) \right >_N =\lambda^N \,
\frac{Z_N (W+[\lambda ])}{Z_N (W)}
$$
where
$$
W+[\lambda ] \, \equiv \,
W(z)+\log \left (1-\frac{z}{\lambda}\right )
$$
is the potential modified
by the (complex and multi-valued) term
$\log \left (1-\frac{z}{\lambda}\right )$.
(Since this term is always under the $\exp$-function,
there is no ambiguity in the choice of its branch.)
In this notation, the
above identities for determinants
acquire the form of the
{\it Hirota bilinear equations} \cite{Hirota}:
\beq\label{Hirota1}
(\lambda_1 \! - \! \lambda_2 ) \, Z_N(W \! +\!
[\lambda_1 ] \! +\! [\lambda_2 ] )\,
Z_N (W\! +\! [\lambda_3  ])\,
 + \, \mbox{cyclic perm-s of $1,2,3$}\, =\, 0
\eeq
\beq\label{Hirota2}
\begin{array}{c}
\! \! \!\!\!\!
Z_N (W) Z_N (W \! +\! [\lambda ]\! +\! [\overline{\lambda} ])
\, - \,
Z_N (W \!+\! [\lambda  ]) Z_N (W \! +\!  [\overline{\lambda} ])
\!\!\!
\\ \\
\displaystyle{=\, \frac{N}{N\! +\! 1}|\lambda |^{-2}
Z_{N+1}(W)\, Z_{N-1}(W \! +\!
[\lambda ] \! +\! [\overline{\lambda} ])}
\end{array}
\eeq
Let us parameterize the potential as
$$
W(z)=W^{(0)}(z) +\sum_k (t_k z^k +\bar t_k \bar z^k)
$$
then these equations state that
$Z_N/N!$, as a function of $t_k$, $\bar t_k$,
is the {\it tau-function} of the 2D Toda lattice
hierarchy.
The transformation $W\rightarrow W+[\lambda]$
is equivalent to the change of variables
$t_k \rightarrow t_k -\frac{1}{k}\lambda^{-k}$
which is known in the literature
as the Miwa transformation.

\paragraph{Orthogonal polynomials and the kernel function.}
The orthogonal polynomials technique is
useful not only for hermitian and unitary
matrix ensembles but for the normal and
complex ensembles as well.
The orthogonal polynomials are introduced
as mean values of the characteristic polynomials
of the random matrices $\Phi$:
\beq\label{orthog1}
P_n(\lambda )=
\lbracket \det (\lambda -\Phi )\rbracket_n
\eeq
Clearly, $P_n$ are polynomials in $\lambda$ of the form
$P_n(\lambda )=\lambda^n +\, \mbox{lower degrees}$.

The main property of the polynomials introduced
is their {\it orthogonality} in the complex plane:
\beq\label{orthog}
\int P_n (z)\overline{P_m(z)}e^{W(z)}d^2 z =
h_n \, \delta_{mn}
\eeq
The square of the norm $h_n =||P_n ||^2$ is
connected with the partition function as
$$
h_n =\frac{1}{n\! +\! 1} \, \frac{Z_{n+1}}{Z_n}\,,
\;\;\;\;\;
Z_N = N! \prod_{n=0}^{N-1}h_n
$$
Again, the proof is completely parallel
to the corresponding proof in the hermitian
models. See Appendix for details.

The functions
$$
\psi_n (z)=\frac{1}{\sqrt{h_{n-1}}}\, P_{n-1}(z) e^{W(z)/2}
$$
are {\it orthonormal}:
\beq\label{orthog2}
\int \psi_n (z)\overline{\psi_m (z)}d^2 z =\delta_{mn}
\eeq
These $\psi_n$'s are ``one-particle wave functions"
of electrons in the magnetic field.
The $N$-particle wave function is
$\Psi_N (z_1 , \ldots , z_N) \sim \det [\psi_{j}(z_k)]$.
The joint probability to find ``particles" at $z_1 ,
\ldots , z_{N}$ is
$$
|\Psi_N (z_1 , \ldots , z_N)|^2 =\frac{1}{N!}\,
|\det [\psi_j (z_k )] |^2
$$
Since $|\det M |^2 = \det (MM^{\dag})$, we can write:
$
|\det [\psi_j (z_k )] |^2 =
\det
\left (
\sum_{n=1}^{N}\psi_n (z_j)\overline{\psi_n (z_k)}
\right )
$.
The expression under the determinant
is called the {\it kernel function}:
\beq\label{kernel}
\kernel_N(z, \bar w)=\sum_{n=1}^{N}
\psi_n (z)\overline{\psi_n (w)}
\eeq
The main properties of the kernel function are:
\begin{itemize}
\item
Hermiticity:
$\kernel_N (z, \bar w)=\overline{
\kernel_N (w, \bar z)}$;
\item
Normalization:
$\int \kernel_N (z, \bar z)d^2 z =N$;
\item
Projection property:
$\int \kernel_N (z_1 , \bar z)\kernel_N (z, \bar z_2)d^2 z
=\kernel_N (z_1 , \bar z_2)$.
\end{itemize}

All density correlation functions can be
expressed through the kernel function.
For example:
$$
\begin{array}{ll}
\lbracket \rho (z)\rbracket_N =
\kernel_N (z, \bar z)&
\\& \\
\displaystyle{
\lbracket \rho (z_1)\rho (z_2) \rbracket_N }=
&\!\! \displaystyle{\left |
\begin{array}{cc}
\kernel_N (z_1 , \bar z_1)&
\kernel_N (z_1 , \bar z_2)
\\&\\
\kernel_N (z_2 , \bar z_1)&
\kernel_N (z_2 , \bar z_2)
\end{array}
\right |
+\,\, \kernel_N (z_1 , \bar z_1 )\delta (z_1 \! - \! z_2 )}
\end{array}
$$
The last term is a {\it contact term}. It does
not contribute if $z_1 \neq z_2$.
In general, one has:
$$
\begin{array}{lll}
\lbracket \rho (z_1 )\ldots \rho (z_n)\rbracket_N
&=&\det (\kernel_N (z_i , \bar z_j))_{1\leq i,j\leq n} +
\, \mbox{contact terms}
\end{array}
$$
where the contact terms vanish if all the points $z_i$
are different.

Note that for the ensemble $\normal$ with an
axially-symmetric potential
$W(\Phi )=W_0 (\Phi \Phi ^{\dag})$
the orthogonal polynomials are simply
$P_n (z)=z^n$ and $h_n$ is given explicitly:
$$
h_n =
\int |z|^{2n} e^{W_0 (|z|^2 )}d^2 z
=2\pi \int_{0}^{\infty}r^{2n+1}
e^{W_0 (r^2)}dr
$$
The kernel function is
$$
\kernel_N (z, \bar w)= e^{\frac{1}{2}(W(z)+W(w))}
\sum_{n=0}^{N-1} \frac{(z\bar w)^n}{h_n}
$$
For example, for the Gaussian model
with $W(z)=-|z|^2$ the squared norms are $h_n = \pi n!$ and
the mean value of density is given by
$$
\lbracket \rho (z) \rbracket_N  =\frac{1}{\pi}
\, e^{-|z|^2}\sum_{n=0}^{N-1}
\frac{|z|^{2n}}{n!}
$$

\paragraph{The Lax representation.}
The orthogonal polynomials obey the recurrence relation of the form
$
zP_n(z)=\sum_{k\leq n}c_{nk}P_k(z)
$.
In terms of the $\psi$-function it reads
$$
z\psi_n (z)=r_n \psi_{n+1}(z)
+\sum_{k\geq 0}u_k (n) \psi_{n-k}(z)
$$
One can represent it as a ``spectral problem"
$
L\psi =z\psi
$
for the difference operator
\beq\label{Lax}
L=r_n e^{\p/ \p n}+\sum_{k\geq 0}
u_k (n) e^{-k\p /\p n}
\eeq
(the Lax operator).
Here $e^{\p /\p n}$ is the shift operator $n \to n+1$ with
the characteristic property $e^{\p / \p n}f(n)=
f(n+1)e^{\p / \p n}$.

Let
$$
W(z)=W^{(0)}(z)+\sum_k (t_k z^k +\bar t_k \bar z^k)
$$
It can be shown that the dependence on the parameters
$t_k$, $\bar t_k$ is given by the
{\it 2D Toda hierarchy}
$$
\frac{\p}{\p t_k} L=[A_k , \, L]\,,
\;\;\;\;
\frac{\p}{\p \bar t_k} L=[L, \, \bar A_k ]
$$
where
$A_k =(L^k )_+ +\frac{1}{2}(L^k)_0$,
$\bar A_k =(L^{\dag k} )_-  +\frac{1}{2}(L^{\dag k})_0$
and
$L^{\dag}=e^{-\p / \p n}r_n +\sum_{k\geq 0}
e^{k\p /\p n}\bar u_{k}(n)$
is the conjugate Lax operator.
Given an operator of the form
$\hat O=\sum_k b_k e^{k \p /\p n}$, we use the standard definition
$(\hat O)_{+}=\sum_{k>0} b_k e^{k \p /\p n}$,
$(\hat O)_{-}=\sum_{k<0} b_k e^{k \p /\p n}$ and
$(\hat O)_{0}=b_0$.
The structure of the Toda hierarchy in models of
random matrices was first revealed in \cite{GMMMO}, see
also review \cite{Morozov}.

In the case of quasiharmonic potential,
it is convenient to modify the $\psi$-functions:
$$
\psi_n \rightarrow \chi_n =\frac{1}{\sqrt{h_{n-1}}}
 e^{V(z)}P_{n-1} (z)
$$
The functions $\chi_n$ obey the orthogonality condition:
$
\int \chi_n (z)  \overline{\chi_m (z)}e^{-|z|^2} d^2 z =\delta_{nm}
$.
Then we have two compatible linear problems:
$$
(L\chi )_n =z \chi_n \,,
\quad \quad
(L^{\dag} \chi )_n =\p_z \chi_n
$$
The second equation can be proven by comparing the matrix elements
of the both sides using
integration by parts.

\subsection{The loop equation}

In matrix models, loop equations are exact relations
which follow from the fact that the matrix integral
defining the model does not depend on changes
of integration variables. In our case, we may start
directly from the integral over eigenvalues (\ref{Z}),
thereby extending the result to any value of $\beta$.

Clearly, the integral (\ref{Z}) remains the same
if we change the integration variables $z_i \rightarrow
\tilde z_i$. In other words,
it is invariant under reparametrizations of the
$z$-coordinate, which we write in the infinitesimal form
as $z_i \rightarrow z_i +\epsilon (z_i)$,
$\bar z_i \rightarrow \bar z_i +
\bar \epsilon (z_i)$.
For the integral
$Z_N =\int e^{-\beta E(z_1 , \ldots , z_N)}
\prod_j d^2 z_j$ with $E$ given in (\ref{energy}),
the reparametrization yields, in the first order:
$$
\prod_j d^2 z_j \longrightarrow
\left [ 1+ \sum_l (\p \epsilon (z_l) \! +\! \bar \p
\bar \epsilon (z_l))\right ]
\prod_j d^2 z_j
$$
$$
E\longrightarrow
E+ \sum_{l}\left ( \frac{\p E}{\p z_l}\, \epsilon (z_i) +
\frac{\p E}{\p \bar z_l}\, \bar \epsilon (z_i)\right )
$$
The invariance of the integral is then expressed by
the identity
$$
\sum_i \int \frac{\p}{\p z_i} \left (
\epsilon (z_i) e^{-\beta E}\right ) \prod_j d^2 z_j =0
$$
valid for any $\epsilon$. Introducing a suitable cutoff
at infinity, if necessary, one sees that
the 2D integral over $z_i$
can be transformed, by virtue of the Green theorem,
into a contour integral around infinity and so it does
vanish.

Let us take
$\epsilon (z_i)=\frac{1}{z-z_i}$, where
$z$ is a complex parameter. The singularity
at the point $z$ does not destroy the above identity
since its contribution is proportional to
the vanishing integral
$\oint d\bar z_i /(z_i -z)$ over a small contour
encircling $z$.
Therefore, we have the equality
$$
\sum_i \int \left [
-\, \frac{\beta \p_{z_i}E}{z-z_i}+\frac{1}{(z-z_i)^2}
\right ] e^{-\beta E} \prod_j d^2 z_j =0
$$
where
$\displaystyle{\p_{z_i}E=-\sum_{l\neq i}\frac{1}{z_i -z_l}
-\beta^{-1} \, \p W(z_i)}$ (see (\ref{energy})).
Using the identity
$$
\sum_{i,j}\frac{1}{(z-z_i)(z-z_j)}
=\sum_{i\neq j}\frac{2}{(z-z_i)(z_i -z_j)}+
\sum_i \frac{1}{(z-z_i)^2}
$$
we rewrite it in the form
$\left < {\cal T}(z_1 , \ldots , z_N)\right >=0$,
where
$$
{\cal T}=
2\sum_i \frac{\p W(z_i)}{z-z_i} \! +\! \beta
\left ( \sum_i \frac{1}{z\! -\! z_i}\right )^2 +
(2\!-\! \beta) \sum_i \frac{1}{(z-z_i)^2}
$$
This identity gives an exact relation between one- and two-point
correlation functions.
To see this, we rewrite it in terms of
$\varphi (z)=-\beta \sum_i \log |z-z_i |^2$
using the rule
$\sum_i f(z_i)=\int f(z)\rho (z)d^2 z$.
The result is
{\it the loop equation}
\beq\label{loopeq}
\frac{1}{2\pi}\int \frac{\p W(\zeta )
\lbracket \Delta \varphi (\zeta )\rbracket}{z-\zeta}
\, d^2 \zeta =\lbracket T(z)\rbracket
\eeq
where
\beq\label{loopeq1}
\phantom{\int}
T(z)=
(\p \varphi (z))^2 +(2\! -\! \beta ) \p^2 \varphi (z)
\phantom{\int}
\eeq
The correlator at coinciding points is understood as
$\displaystyle{\lbracket (\p \varphi (z))^2 \rbracket
=\lim_{z'\to z}
\lbracket \p \varphi (z) \, \p \varphi (z')\rbracket}$.

We have got an
{\it exact} relation
between one- and two-point correlation functions,
valid for any finite $N$. For historical reasons, it is
called the loop equation.
One may read it as a Ward identity obeyed by correlation
functions of the model.
Being written in the form (\ref{loopeq}), (\ref{loopeq1}), it
resembles conformal Ward identities.
Since correlation functions are
variational derivatives of the free energy, the loop equation
is an implicit functional relation for the free energy.
However, it is not a closed relation.
It can be made closed by some additional assumptions
or approximations.
A combination with $1/N$ expansion
is particularly meaningful.

\section{Large $N$ limit}

Starting from this section, we study the
large $N$ limit of the
random matrix models introduced in Section 2.
Our main tool is the loop equation.
We shall see that in the large $N$ limit
meaningful geometric and algebro-geometric structures
emerge, as well as important applications in physics.

\subsection{Preliminaries}

In order to be prepared for taking the
large $N$ (``quasiclassical") limit,
it is convenient to introduce the ``Planck constant" $\hbar$
by the rescaling
$W(z) \to \frac{1}{\hbar}W(z)$,
so the integral for the partition function acquires
the form
\beq\label{Zwithhbar}
Z_N =
\int [D\Phi ]  e^{\frac{1}{\hbar}\mbox{tr}\, W(\Phi )}\propto
\int |\Delta_N |^{2\beta} \prod_j
e^{\frac{1}{\hbar}W(z_j)}d^2 z_j
\eeq
which is ready for an $\hbar$-expansion.

Now we can specify what we mean
by the large $N$ limit.
Namely, we are going to consider
the integral (\ref{Zwithhbar}) in the limit
$$
N\to \infty ,  \quad
\hbar \to 0, \quad
\hbar N =t \quad \mbox{finite},
$$
where $t$ is a (positive)
parameter having the dimension of area,
or, equivalently, the integral
$$
Z_N = \int |\Delta_N (z_i)|^{2\beta}
\prod_j
\exp \left (\frac{N}{t} W(z_j)\right ) d^2 z_j
$$
as $N\to \infty$. With this convention, the $N\to \infty$
and $\hbar \to 0$ limits mean the same.
It is natural to expect that
$Z_N \stackrel{\hbar \to 0}{\longrightarrow} e^{\hbar^{-2} F_0 (t)}$,
where the rescaled free energy $F_0$
is a smooth function of $t$ (and parameters
of the potential) finite as $N\to \infty$.

From now on we {\it change the normalization} of
the functions $\rho$ and $\varphi$
(see (\ref{density1}), (\ref{potential1}))
multiplying them by the $\hbar$:
\beq\label{functions1}
\rho (z)=\hbar \sum_i \delta (z-z_i)\,,
\;\;\;\;\;
\varphi (z)=-\hbar \beta \sum_i \log |z-z_i |^2
\eeq
and use these definitions hereafter. (The idea is
to make their mean values finite as $N\to \infty$.)
The relation (\ref{rhophi}) between these functions
remains unchanged.
The density is now normalized as follows:
$$
\int \rho (z)d^2 z =t
$$
Note that in our units the $\hbar$ has dimension
of $[\mbox{length}]^2$, $\rho (z)$ is dimensionless and
the partition function defined by (\ref{Zwithhbar})
has dimension  $[\mbox{length}]^{N(\beta N +2 -\beta )}$,
i.e., the combination
\beq\label{dimensionless}
\hbar^{-\frac{1}{2}\beta N^2 +\frac{1}{2}(\beta -2 )N} \, Z_N
=\int \left | \Delta_N \left (z_i / \sqrt{\hbar}\right )
\right |^{2\beta}
\prod_{j=1}^{N} \left (
e^{\frac{1}{\hbar} W(z_j)} \, \frac{d^2 z_j}{\hbar} \right )
\eeq
is dimensionless.

In terms of the renormalized $\varphi$, the loop equation
(\ref{loopeq}) has the form
\beq\label{loopeq1a}
\frac{1}{2\pi}\int \frac{\p W(\zeta )
\lbracket \Delta \varphi (\zeta )\rbracket}{\zeta -z}
\, d^2 \zeta +
\lbracket (\p \varphi (z))^2 \rbracket
 +\varepsilon \lbracket \p^2 \varphi (z)\rbracket
=0
\eeq
where
\beq\label{varepsilon}
\varepsilon = (2-\beta ) \hbar
\eeq
is small as $\hbar \to 0$. Note that $\varepsilon$ is exactly
zero for the ensemble $\normal^0$ ($\beta =2$), and so
the last term does not enter the loop equation in this case.
It is convenient to treat $\varepsilon$ and $\hbar$
as independent small parameters.

\subsection{Solution to the loop equation in the leading order}

It is instructive to think about the large $N$ limit
under consideration
in terms of the Dyson gas picture.
Then the limit we are interested in corresponds to
a very low temperature of the gas, when fluctuations
around equilibrium positions of the charges are
negligible. The main contribution to the partition function
then comes from a configuration,
where the charges are ``frozen" at their equilibrium
positions. It is also important that
the temperature tends to zero simultaneously with increasing the
number of charges, so the plasma can be regarded as a
continuous fluid at static equilibrium.
In the noninteracting fermions picture, this limit has
some features of the quasiclassical approximation.
Mathematically, all this means that the integral is
evaluated by the saddle point method, with only the
leading contribution being taken into account.
As $\hbar \to 0$, correlation functions
take their ``classical" values
$\lbracket \varphi (z) \rbracket = \varphi _{cl}(z)$,
and multipoint correlators factorize in the
leading order:
$\lbracket \p \varphi (z) \,
\p \varphi (z')\rbracket =
\p \vphcl (z) \p\vphcl (z')$.
Then the loop equation (\ref{loopeq1a})
becomes a {\it closed} relation for $\vphcl$:
\beq\label{loopeq2}
\frac{1}{2\pi} \int \frac{\p W( \zeta )\Delta
\varphi _{cl}(\zeta )}{\zeta -z}\, d^2 \zeta \quad
+\, \Bigl ( \p \vphcl (z)\Bigr )^2 +
\varepsilon \,
\p^2 \vphcl (z)\, = \, 0
\eeq
Note that we hold the last term which is apparently
of the next order in $\hbar$. The role of this term
will be discussed below.

\paragraph{The case $\beta =2$ (the ensemble $\normal^0$).}
We begin with the case $\beta =2$, when the last term in the r.h.s.
of (\ref{loopeq2}) vanishes exactly.
Let us apply $\bar \p$ to both sides of the equation.
This yields:
$$
-\p W(z) \Delta \vphcl (z)+
\p \vphcl (z) \Delta \vphcl (z)=0
$$
Since $\Delta \vphcl (z) \propto \rhocl (z)$
(see (\ref{functions1})),
we obtain
\beq\label{large1}
\rhocl (z) \, \left [ \p \vphcl (z) -\p W(z)\right ]=0
\eeq
This equation should be solved with the
additional constraints
$\int \rhocl (z) d^2 z =t$ (normalization) and
$\rhocl (z)\geq 0$ (positivity).
The equation tells us that
$\mbox{either} \quad \p \vphcl (z)=\p W(z)$ or
$\rhocl (z)=0$.
Applying $\bar \p$,
we get $\Delta \vphcl (z)=\Delta W(z)$.
Since $\Delta \varphi =-4\pi \beta \rho$,
this gives the solution for $\rho_{cl}$:
\beq\label{large2}
\rhocl (z)=-\, \frac{\Delta W(z)}{4\pi \beta}
\quad \quad \mbox{``in the bulk"}
\eeq
Here, ``in the bulk" just means ``in the region where
$\rhocl > 0$". As we shall
see below, this result holds true, up to
some details, for other values of
$\beta$ as well, so $\beta$ is kept
in this formula and in some formulas below.

The physical meaning of the equation
$\p \vphcl (z) =\p W(z)$
is clear. It is just the condition that
the charges are in equilibrium (the saddle point for the
integral).
Indeed, the equation states that the total force
experienced by a charge at any point $z$, where
$\rhocl \neq 0$,
is zero. The interaction with the other charges,
$\p \vphcl (z)$, is compensated by the
force $\p W(z)$ due to the external field.

\begin{figure}[tb]
\epsfysize=3cm
\centerline{\epsfbox{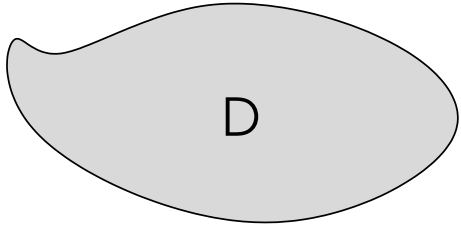}}
\caption{\sl The support of eigenvalues.}
\label{fi:support}
\end{figure}

\paragraph{Support of eigenvalues.}
Let us assume that
\beq\label{sigma}
\sigma (z):=-\frac{1}{4\pi}\Delta W(z) >0
\eeq
For quasiharmonic potentials, $\sigma (z)=1/\pi$.
If, according to (\ref{large2}),
$\rhocl =\sigma /\beta$ everywhere, the normalization
condition for $\rhocl$ can not be satisfied!
So we conclude that
$\rhocl = \sigma /\beta$ in a compact bounded domain
(or domains) only, and outside this domain one should
switch to the other solution of (\ref{large1}),
$\rhocl =0$.
The domain ${\sf D}$ where $\rhocl >0$ is called
{\it support of eigenvalues} (Fig.~\ref{fi:support}).
In general, it may
consist of several disconnected components.
The {\it complement} to the support of eigenvalues,
${\sf D^c} = {\bf C}\setminus {\sf D}$, is an unbounded
domain in the complex plane.
For quasiharmonic potentials, the result is especially
simple: $\rhocl$ is constant in ${\sf D}$ and
$0$ in ${\sf D^c}$.

In terms of the mean value of the function $\varphi (z)$
in the leading order, the above result reads
$$
\vphcl (z)=-\int_{{\sf D}}
\log |z-\zeta |^2 \sigma (\zeta )\, d^2 \zeta
$$
As it follows from the theory of potential in two
dimensions, this function is continuous across the
boundary of ${\sf D}$ together with its first
derivatives. However, the second order derivatives
of this function have a jump across the boundary.

To find the shape of ${\sf D}$ is a much more
challenging problem. It appears to be equivalent to the inverse
potential problem in two dimensions.
The shape of ${\sf D}$ is determined by the condition
$\p \vphcl (z) =\p W(z)$ (imposed for all points $z$
inside ${\sf D}$)
and by the normalization condition. Since
$\p \vphcl (z) = -\beta \int \frac{\rhocl (\zeta )
d^2 \zeta}{z-\zeta}$ (see (\ref{functions1})), we write
them in the form
$$
\left \{
\begin{array}{l}
\displaystyle{
\frac{1}{4\pi} \int_{{\sf D}}
\frac{\Delta W(\zeta )d^2 \zeta}{z-\zeta} \, = \,
\p W(z)} \quad \quad \mbox{for all $z \in {\sf D}$}
\\  \\
\displaystyle{
\int_{{\sf D}}
\sigma (\zeta ) d^2 \zeta =\beta t}
\end{array}
\right.
$$
The integral
over ${\sf D}$ in the first equation
can be transformed to a contour integral
by means of the Cauchy formula (see Appendix B).
As a result, one obtains:
\beq\label{large3}
\oint_{\p {\sf D}} \frac{\p W(\zeta )d\zeta}{z-\zeta}
=0
\quad \quad \mbox{for all $z\in {\sf D}$.}
\eeq
This means that the domain ${\sf D}$
has the following property:
the function $\p W(z)$
on its boundary is the boundary value of an analytic
function in its complement ${\sf D^c}$.

We continue our analysis
for the quasiharmonic case, where
$$
\p W(z)=-\bar z +V'(z)\,,
\quad
\Delta W(z)=4\p\bar \p W(z)=-4
$$
The normalization than means that the area of ${\sf D}$
is equal to $\beta \pi t$.
Assume that:
\begin{itemize}
\item[-]
$V(z)=\sum t_k z^k$ is regular in ${\sf D}$
(say a polynomial)
\item[-]
$0\in {\sf D}$ (it is always the case if $-W$ has a local
minimum at $0$)
\item[-]
${\sf D}$ is connected
\end{itemize}
Then the first equation in (\ref{large3})
acquires the form
$$
\frac{1}{2\pi i}\oint_{\p {\sf D}}
\frac{\bar \zeta d \zeta}{\zeta -z}=V'(z)
\quad \quad \mbox{for $z \in {\sf D}$}
$$
Expanding it near $z=0$, we get:
\beq\label{harmmom}
t_k =\frac{1}{2\pi i k}
\oint_{\p {\sf D}} \bar \zeta \zeta ^{-k} d\zeta =
-\frac{1}{\pi k} \int_{{\sf D}} \zeta ^{-k}d^2 \zeta
\eeq
We see that the ``coupling constants" $t_k$
are {\it harmonic moments}
of ${\sf D^c}={\bf C}\setminus {\sf D}$
and the area of ${\sf D}$ is $\pi \beta t$.

It is the subject of the inverse potential problem
to reconstruct the domain from its
area and harmonic moments.
(In the case when the support of
eigenvalues has several disconnected
components, some additional conditions are required.)
In general, the problem has many solutions.
But it is known that {\it locally}, i.e.,
for a small enough change $t \rightarrow t+\delta t$,
$t_k \, \rightarrow \, t_k +\delta t_k$
the solution is unique.

\paragraph{The support of eigenvalues: a fine structure
($\beta \neq 2$).}
Let us take a closer look at the support of eigenvalues,
taking into account the so far ignored term in (\ref{loopeq2}).
Applying $\bar \p$ to the both sides of (\ref{loopeq2})
yields,
instead of (\ref{large1}), the equation
$$
\p \vphcl (z)-\p W(z)\, +
\frac{\varepsilon}{2}\,
\p \log \Delta \vphcl (z) =0
$$
with the extra term proportional to
$\varepsilon$ (\ref{varepsilon}).
Acting by $\bar \p$ once again,
we obtain
the equation for $\rhocl$
\beq\label{liouville}
-\, \frac{\varepsilon}{8\pi}
\Delta \log \rhocl (z) +\beta \rhocl (z)=\sigma (z)
\eeq
which looks like the Liouville equation in the ``background"
$\sigma (z)$.
The first term seems to be negligible as $\hbar \to 0$.
However, one should be careful
since the small parameter stands in front of the term with
the highest derivative. As a matter of fact, this term
is negligible only if $\rhocl$ is not small!
Indeed, in a region where
$\rhocl \sim e^{-N}$, the term
$\varepsilon \Delta \log \rhocl$ is of order 1 and
thus plays the dominant role.

In fact the equation states that $\rho_{cl}$
never vanishes exactly but can be exponentially
small as $N\to \infty$. In the bulk, where the charges
are distributed with nonzero density at $N=\infty$,
equation (\ref{liouville})
systematically generates corrections
to the value $\sigma (z)/\beta$. If $\sigma \neq
\mbox{const}$, there are power-like corrections
in $\varepsilon$, as well as exponentially small ones.
We conclude that the effect of the
extra term is negligible everywhere except
the very vicinity of the edge of the support
of eigenvalues.
Therefore, the result for $\rhocl$ can still
be written in the form
\beq\label{rhocl}
\rhocl (z)= \beta^{-1}\sigma (z) \, \Theta (z; {\sf D})
\eeq
where $\Theta (z; {\sf D})$ is the characteristic function
of the domain ${\sf D}$ (which is $1$ in ${\sf D}$ and
$0$ in ${\sf D^c}$).
The role of the term $\varepsilon \Delta \log \rhocl$
is to make
the edge smooth.
Around the edge,
the density rapidly (but smoothly)
drops down to zero over distances of
order $\sqrt{\varepsilon}$.
So, the boundary has got
a ``fine structure".

All this is in agreement with the form of the
first nonvanishing correction to the mean
density found from the loop equation (\ref{loopeq1a}).
Let us write $\lbracket \varphi (z)\rbracket =
\vphcl (z)+\varphi_{\hbar }(z)$, where
$\varphi_{\hbar }$ is of order $\hbar$.
The result for the $\varphi_{\hbar }$
can be compactly written in terms of the function
\beq\label{chi}
\chi (z)=\log \sqrt{\pi \sigma (z)}
\eeq
and its harmonic continuation $\chi^H (z)$
from the boundary of
${\sf D}$ to its exterior.
Specifically, $\chi^H (z)$ is a harmonic function in
${\sf D^c}$ (regular at $\infty$) such that
$\chi^H (z) = \chi (z)$ on the boundary.
In other words, it is the solution of the (exterior)
Dirichlet boundary value problem (see below).
The loop equation yields
\beq\label{phihbar}
\varphi_{\hbar}(z)=
\left \{
\begin{array}{ll}
-\varepsilon \left [ \chi (z) -
\chi^H (\infty )+\frac{1}{2}\right ],
& z\in {\sf D}
\\ &\\
-\varepsilon \left [ \chi^H (z) -
\chi^H (\infty ) \, \right ],
& z\in {\sf D^c}
\end{array}
\right.
\eeq
(Note the discontinuity of this function across the
boundary.)
The corresponding correction to the mean density
$\rho_{\hbar }= \lbracket \rho \rbracket
-\rhocl$ is
\beq\label{rho1/2}
\rho_{\hbar}(z)=
\frac{\varepsilon}{4\pi \beta}\left (
\Theta (z; {\sf D}) \Delta \chi (z) -
\delta (z; \p {\sf D}) \p_n \left (\chi (z)- \chi^H (z)\right )
-\frac{1}{2} \delta ' (z; \p {\sf D})\right )
\eeq
where $\delta (z; \p {\sf D})$ is the delta function
with the support on the boundary and
$\delta ' (z; \p {\sf D})$ is its normal derivative
(see Appendix B).
Here and below, $\p_n$
is the normal derivative at the
boundary, with the
normal vector being directed to the
exterior of the domain ${\sf D}$.
The correction is so singular
because the zeroth approximation (\ref{rhocl})
is singular by itself.
The singular function $\rho_{\hbar}$ is to be understood as
being integrated with any smooth test function.
The first term in (\ref{rho1/2})
is a correction to the bulk density. The second one
is a correction to the shape of the support of eigenvalues
(it describes a small displacement of the edge).
The third term signifies the presence of a double layer of charges
around the boundary. This just means that the boundary is
smoothed out.

\begin{figure}[tb]
\epsfysize=6.5cm
\centerline{\epsfbox{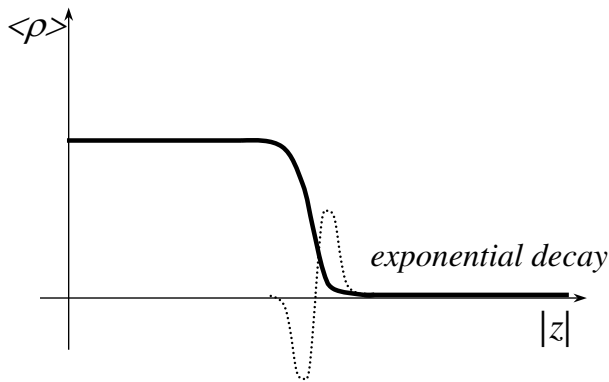}}
\caption{\sl The mean density profile
for models with quasiharmonic potential
in the large $N$ limit.
The correction to the pure step function
is shown by the dotted line. The effect
of the correction is to form a double layer
of charges near the edge.}
\label{fi:profile}
\end{figure}

For the normal self-dual matrices
($\beta =2$) the correction
$\rho_{\hbar}$ vanishes. Certainly, this does not mean that
the boundary is sharp. It becomes smooth if
higher corrections in $\hbar$ (caused by
fluctuations of the particles) are taken into account.

For the normal matrix model ($\beta =1$) with
quasiharmonic potential equation (\ref{liouville})
reads
$$
-\, \frac{\hbar}{8}
\Delta \log \rhocl (z) +\pi \rhocl (z)=1
$$
The obvious solution is $\rhocl =1/\pi$.
But it is not normalizable!
One must look for another solution.
The right solution differs from the constant
by exponentially small terms in the bulk but
exhibits an abrupt drop across the boundary
of the domain determined by the harmonic moments
(\ref{harmmom}).
So, up to exponentially small corrections,
the solution is given by (\ref{rhocl}) with $\sigma =1/\pi$
and $\beta =1$. It is instructive to note
that the first two terms in (\ref{rho1/2}) vanish
at $\sigma =\mbox{const}$ but the third term
does not, so the leading correction in the quasiharmonic
case merely makes the edge smooth that results in
a double layer of ``charges" near the edge
(see Fig.~\ref{fi:profile}).

\paragraph{From the support of eigenvalues to an
algebraic curve.}
There is an interesting algebraic geometry behind the
large $N$ limit of matrix models.
For simplicity, here we
consider models with quasiharmonic potentials.

In general, the boundary of the support of eigenvalues
is a closed curve in the plane without self-intersections.
The following important fact holds true.
If $V'(z)$ is a rational function, then
this curve
is a real section of a complex
algebraic curve of finite genus.
In fact, this curve encodes the
$1/N$ expansion of the model.
In the context of Hermitian 2-matrix model
such a curve was introduced and studied
in \cite{Staud,KM,Eyn-curve}.

To explain how the curve comes into play, we start from
the equation $\p \vphcl = \p W$, which
can be written in the form
$\bar z -V'(z)=G(z)$ for $z\in {\sf D}$, where
$$
G(z)=\frac{1}{\pi}\int_{{\sf D}}
\frac{d^2 \zeta}{z-\zeta}
$$
Clearly, this function is analytic in ${\sf D^c}$.
At the same time,
$V'(z)$ is analytic in ${\sf D}$ and all its
singularities in
${\sf D^c}$ are poles.
Set
$$
S(z)=V'(z)+G(z)
$$
Then
$
S(z)=\bar z
$
on the boundary of the support of eigenvalues.
So, $S(z)$ is the analytic continuation of $\bar z$
away from the boundary.
Assuming that poles of $V'$
are not too close to $\p {\sf D}$, $S(z)$ is
well-defined at least
in a piece of ${\sf D^c}$ adjacent to the boundary.
The complex conjugation yields $\overline{S(z)}=z$,
so the function $\bar S(z)=\overline{S(\bar z)}$ must
be inverse to the $S(z)$:
$$
\bar S(S(z))=z
$$
(``unitarity condition").
The function
$S(z)$ is called the {\it Schwarz function} \cite{Davis}.

Under our assumptions, the $S(z)$ is an algebraic function,
i.e., it obeys a polynomial equation
$R(z, S(z))=0$ of the form
$$
R(z, S(z))=\sum_{n,l=1}^{d+1} a_{nl} z^n (S(z))^l =0
$$
where $\overline{a_{ln}}= a_{nl}$ and
$d$ is the number of poles of $V'(z)$ (counted
with their multiplicities).
Here is the sketch of proof.
Consider the Riemann surface
$\mit\Sigma ={\sf D^c} \cup \p {\sf D} \cup ({\sf D^c})^*$
(the {\it Schottky double} of ${\sf D^c}$).
Here, $({\sf D^c})^*$
is another copy of ${\sf D^c}$, with the local coordinate $\bar z$,
attached to it along the boundary. On $\mit\Sigma$, there
exists an anti-holomorphic involution that interchanges
the two copies of ${\sf D^c}$ leaving the points of $\p {\sf D}$
fixed.
The functions $z$ and
$S(z)$ are analytically extendable to
$({\sf D^c})^*$ as $\overline{S(z)}$ and $\bar z$ respectively.
We have two meromorphic functions, each with $d+1$ poles,
on a closed Riemann surface. Therefore, they are connected by
a polynomial equation of degree $d+1$ in each variable.
Hermiticity of the coefficients follows from the unitarity condition.

The polynomial equation $R(z, \tilde z)=0$
defines a complex curve
$\mit\Gamma$
with anti-holomorphic involution
$(z, \tilde z)\mapsto
(\overline{\tilde z}, \bar z)$.
The real section is the set of points such that
$\tilde z = \bar z$.
It is the boundary of the support
of eigenvalues.

It is important to note that for models
with non-Gaussian weights (in particular,
with polynomial potentials of degree greater
than two) the curve has
a number of singular points, although the Riemann
surface $\mit\Sigma$ (the Schottky double) is smooth.
Generically, these are
{\it double points}, i.e., the points where the curve
crosses itself. In our case, a double point
is a point $z^{(d)}\in {\sf D^c}$
such that $S(z^{(d)})=\overline{z^{(d)}}$ but $z^{(d)}$
does not belong to the boundary of ${\sf D}$. Indeed,
this condition means that two different points on $\mit\Sigma$,
connected by the antiholomorphic involution,
are stuck together on the curve $\mit\Gamma$, which means
the self-intersection.
The double points play the key role
in deriving the nonperturbative (instanton) corrections to
the large $N$ matrix models results (see \cite{nonperturbative}
for details).

Finally, let us point out that
a complex curve $\mit\Gamma^{(n)}$ can be associated to
ensembles of finite matrices as well
(at least for $\beta =1$). For the model of
two Hermitian matrices this was done in \cite{BEH}.
If the linear spectral problem for the $L$-operator
is of finite order
(see the end of Section 3.2), the curve can be
defined as the
``spectral curve" of the difference spectral problems
$(L\chi )_n =z\chi_n$,
$(L^{\dag}\chi )_n = \tilde z \chi_n$.
Since the operators $L$ and $L^{\dag}$ do not commute,
the curve depends on $n$. (See \cite{teodor}
for details.) In contrast to the curve
$\mit\Gamma$ the curve $\mit\Gamma^{(n)}$
is in general a smooth curve.
A properly performed $n\to \infty$ limit of this curve
coincides with the complex curve $\mit\Gamma$ constructed from
the support of eigenvalues.

\subsection{The free energy}

In this subsection we assume that the support of
eigenvalues is connected. As is known,
the free energy admits a $1/N$
expansion. We prefer to work with the equivalent
$\hbar$-expansion,
thus emphasizing its semiclassical nature.
The first few terms of the $\hbar$-expansion
for the ensembles of normal matrices
with a general potential are
\beq\label{hbarexp}
\log Z_N = c (N) + \frac{F_0}{\hbar^2} +
\frac{F_{1/2}}{\hbar} + F_1 + O(\hbar)
\eeq
The explicit form of the
$c(N)$ is given below.
In fact this term
can be absorbed into a normalization.
For example,
one may normalize $Z_N$ dividing it
by the partition function of the Gaussian model.
Although in general the $\hbar$-expansion
does not look like a topological one,
it appears to be
topological (i.e., only even powers of $\hbar$ enter)
for the ensemble $\normal^0$ with arbitrary potential and
for the ensemble $\normal$ with quasiharmonic potential.

\paragraph{The leading order.}
The partition function is given by (\ref{ZE}),
$Z_N = \int e^{-\beta E}\prod d^2 z_j$, where
\beq\label{betaE}
-\beta E(z_1 , \ldots , z_N)=
\beta \sum_{i\neq j} \log |z_i - z_j | + \hbar^{-1}
\sum_j W(z_j)
\eeq
Writing the energy in terms of the density function, we have,
in the leading order:
\beq\label{betaE1}
-\beta \hbar^2 E [\rho ]=\beta
\int\!\!\int \rho (z)\rho
(\zeta )\log |z-\zeta | d^2 z d^2 \zeta
\, +\, \int W(z)\rho (z) d^2 z
\eeq
We need to find the minimum of $E[\rho ]$ with the
constraint $\int \rho \, d^2 z =t$.
This is achieved by variation
of the functional
$E [\rho ] +\lambda (\int \rho \, d^2 z -t)$ with the
Lagrange multiplier $\lambda$. The resulting
equation is
$$
2\beta \int \log |z-\zeta | \rho (\zeta ) d^2 \zeta + W(z) +\lambda =0
$$
Upon taking the $z$-derivative,
we see that
the extremal $\rho (z)$
is equal to the $\rhocl (z)$, as expected, and the
equation coincides with
the previously derived one,
$\p \vphcl (z) =\p W(z)$, with
$$
\vphcl (z)= -\beta \int \log |z-\zeta |^2 \rhocl (\zeta )
d^2 \zeta =
 -\int_{{\sf D}} \log |z-\zeta |^2 \sigma (\zeta )
d^2 \zeta
$$
Assuming that $W(0)=0$
and ${\sf D}$ is connected, the Lagrange multiplier
is fixed to be $\lambda = \vphcl (0)$, and so
$W(z)= \vphcl (z)- \vphcl (0)$.
Plugging this into (\ref{betaE1}),
we find the leading contribution to the free energy
$F_0 /\hbar^2 = \mbox{max}_{\rho}
(-\beta E[\rho ])=-\beta E[ \rhocl ]$:
\beq\label{free1}
F_0 = - \frac{1}{\beta}\int_{{\sf D}}\!\! \int_{{\sf D}}
\sigma (z)\log \left | \frac{1}{z}\!-\! \frac{1}{\zeta}\right |
 \sigma (\zeta) d^2 z d^2 \zeta
\eeq
which is basically the electrostatic energy
of the domain ${\sf D}$ charged with the density $\sigma (z)$
with a point-like compensating charge at the origin.

Since the $t$-derivative of the extremal value of the
functional is equal to the Lagrange multiplier (with the sign minus),
$\p_t F_0 = -\lambda$,
we incidentally obtain the useful formula
\beq\label{gr3a}
\p_t F_0  = 2\int_{{\sf D}} \log |z| \, \sigma (z)
\, d^2 z
\eeq
which will be rederived below by
a more direct method.

\paragraph{Corrections to the leading term.}
Taking into account
the discrete ``atomic" structure of the Dyson gas,
one is able to find the subleading corrections to
the free energy.

The first correction comes from a more accurate
integral representation of the sum
$\sum_{i\neq j} \log |z_i - z_j |$, when
passing to the continuous theory. Namely, one should
exclude the terms with $i=j$, writing
$$
\sum_{i\neq j} \log |z_i - z_j | =
\sum_{i,j} \log |z_i - z_j |
-\sum_j \log |\ell (z_j )|
$$
where $\ell$ is a short-distance cutoff
(which may depend on the point $z_j$). It is natural to
take the cutoff to be
\beq\label{cutoff}
\ell (z) \sim \sqrt{ \frac{\hbar}{\rhocl (z)}}
\eeq
which is the mean distance between
the charges around the point $z$.
(In the context of the quantum Hall effect,
$\ell \sim \sqrt{\hbar / B}$ is called the magnetic
length.)
This gives the
improved estimate for $E[\rhocl ]$:
\beq\label{smdist}
-\beta \hbar^2 E[\rhocl ] = F_0 + \beta \hbar
\int \rhocl (z) \sqrt{\rhocl (z)} \, d^2 z -
\frac{1}{2}\beta N \log \hbar
+\alpha_1 N
\eeq
where $\alpha_1$ is a numerical constant which can not be determined
by this argument.

Another correction comes from the integration measure
when one passes from the integration over $z_j$ to the
integration over macroscopic densities\footnote{I thank
A.Abanov for a discussion on this point.}. We can write
$$
\prod_j d^2 z_j = N! \, J[\rho ] \, [D\rho ]
$$
where $[D\rho ]$ is an integration measure in the space
of densities, $J[\rho ]$ is the Jacobian of this change of variables
and the factor $N!$ takes into account
the symmetry under permutations (all the states
that differ by a permutation of the charges
are identical). To estimate the Jacobian,
we divide the plane into $N$ microscopic ``cells" such that
$j$-th particle occupies a cell of size
$\ell (z_j)$, where
$\ell (z_j)$ is the mean distance (\ref{cutoff})
between the particles
around the point $z_j$.
All the microscopic states in which the particles
remain in their cells are macroscopically indistinguishable.
Given a macroscopic density $\rho$,
$J[\rho ]$ is then approximately equal to the integral
$\int_{{\rm cells}} \prod_j d^2 z_j$, with each particle
being confined to
its own cell. Therefore,
$J[\rho ] \sim \prod_j \ell^2 (z_j)$, and thus
$\log J[\rho ]$ (sometimes referred to as entropy
of the state with the macroscopic density $\rho$)
is given by
\beq\label{Jacob}
\log J[\rho ] = -\frac{1}{\hbar} \int \rhocl (z)
\log \rhocl (z) \, d^2 z \, + \, N \log \hbar +\alpha_2 N
\eeq
where $\alpha_2$ is a numerical constant.
This result agrees with the corresponding Jacobian
obtained within the
collective field theory approach \cite{collective}.

Combining (\ref{smdist}), (\ref{Jacob}) with
$\rho = \rhocl$, and taking into account the factor $N!$
in the measure, we obtain:
\beq\label{cN}
c(N)= \log N! + \frac{N}{2}(2-\beta ) \log \hbar
+\alpha N
\eeq
where $\alpha$ is a numerical constant, and
\beq\label{F1/2}
F_{1/2} = -\, \frac{2-\beta}{2\beta}\int_{{\sf D}}
\sigma (z) \log (\pi \sigma (z)) \, d^2 z
\eeq
The term $F_{1/2}$ is thus the sum of the contribution
due to the short-distance cutoff and the entropy contribution,
which cancel each other in the ensemble of normal self-dual
matrices (at $\beta =2$).
Another remarkable case
when $F_{1/2}$ vanishes exactly\footnote{However,
its variational derivative, $\rho_{\hbar}$,
does not vanish, as is seen from (\ref{rho1/2}).} is the case of
quasiharmonic potentials.

The result for $F_{1/2}$ can be derived in a more
rigorous way from the loop equation (see Appendix C).
Here we simply note
that the variation of (\ref{F1/2}) over the potential $W$
does yield the correction to the mean density given by
(\ref{rho1/2}). One can verify this using the
variational technique presented below in Section 4.4.

The result for $c(N)$ is in agreement with the dimensionality
argument. It is easy to see that $e^{F_0/\hbar^2}$ carries
the dimension of $[\mbox{length}]^{\beta N^2}$, and higher
terms are dimensionless. The dimension of $Z_N$ is given by
(\ref{dimensionless}). Therefore,
$e^{c(N)}$ must carry the residual dimension
$[\mbox{length}]^{N(2-\beta )}$, which agrees
with (\ref{cN}). For quasiharmonic potentials, the constant
$\alpha$ can be found explicitly:
$\alpha = \log\sqrt{2\pi^3}$,
as is readily seen from the
Gaussian case.

To summarize, the asymptotic expansion
of the partition function as $\hbar \to 0$
has the form
\beq\label{hbarexp1}
Z_N = N! \hbar^{\frac{1}{2}(2-\beta )N} e^{\alpha N}
\exp \left ( \frac{F_0}{\hbar^2} +
\frac{F_{1/2}}{\hbar} +F_1 +\sum_{k\geq 3} \hbar^{k-2}F_{k/2}
\right )
\eeq
where $F_0$ and $F_{1/2}$ are given by (\ref{free1})
and (\ref{F1/2}) respectively. The higher corrections
are due to fluctuations of the eigenvalues around the
equilibrium configuration. No simple method to
find their explicit form is known. In principle,
these corrections can be found by expanding the loop
equation in powers of $\hbar$ (see Appendix C),
similarly to how it goes for the model of one Hermitian
matrix \cite{F1H,F1multi}.
However, the calculations are rather tedious, even in the
first two orders. At present
only fragmentary results are available.
Some of them look quite suggestive.

For example, the result for the
$F_1$-correction obtained in \cite{WZF1} for the case
of the normal matrix model with quasiharmonic potential
and a connected support of eigenvalues
has a clean interpretation as the free energy of the
theory of free bosons in the domain ${\sf D^c}$ with
the Dirichlet boundary conditions. Namely, the
calculations yield the result
\beq\label{F1quasi}
F_1  =
-\,\, \frac{1}{24 \pi} \oint_{|w|=1}
\Bigl (\log |z'(w)|\p_n \log |z'(w)| +
2\log |z'(w)| \Bigr )|dw|
\eeq
where $z(w)$ is the univalent conformal map from
${\sf D^c}$ onto the exterior of the unit circle.
Comparison with the Polyakov-Alvarez formula
\cite{Polyakov,Alvarez}
allows one to identify this quantity with
$-\frac{1}{2}\log \det (-\Delta_{{\sf D^c}})$, where
$\det (-\Delta_{{\sf D^c}})$ is
the regularized determinant of the Laplace operator
in ${\sf D^c}$ with the Dirichlet boundary conditions.
This suggests the interpretation through free
bosons\footnote{A similar interpretation of the
$F_1$-correction in the Hermitian matrix ensemble
was suggested in \cite{Kostov}.}.
Presumably, the higher corrections to the free energy
are connected with spectral geometry of the Laplace
operator, too.
Recently, some progress in computation of $F_1$
and, more generally, in understanding the structure
of the whole series (including the case of
disconnected supports)
in models of Hermitian matrices was achieved
\cite{F1}. Conjecturally, the answer is to be
expressed in terms of a (conformal?) field theory on the
complex curve $\mit \Gamma$ introduced at the end
of Section 4.2.

Finally, we note that the structure of the loop equation
suggests to rearrange the $\hbar$-expansion of the free energy
and to write it in the ``topological" form
$F= \sum_{g\geq 0} \hbar^{2g} F_g$, where each term has its
own expansion in  $\varepsilon =(2-\beta )\hbar$:
$F_g = F_{g}^{0}+\sum_{n\geq 1} \varepsilon^n F_{g}^{(n)}$.

\subsection{Correlation functions in the large $N$ limit}

\begin{figure}[tb]
\epsfysize=3.7cm
\centerline{\epsfbox{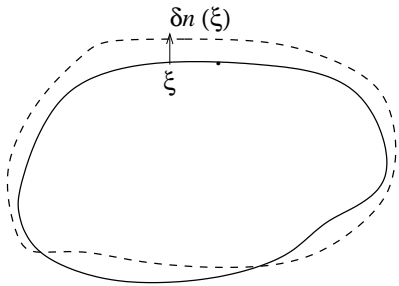}} \caption{\sl The
normal displacement of the boundary. } \label{fi:var}
\end{figure}

\paragraph{Variational technique and the Dirichlet
boundary value problem.}
Correlation functions in the leading order in $\hbar$
can be obtained from the free energy
by variation w.r.t. $W(z)$
according to the formulas from Section 3.1.
For a variation of the
potential,
$W\rightarrow W+ \delta W$ with $N\hbar$ fixed
we ask how ${\sf D}$ changes.
It is convenient to describe small deformations
${\sf D}\rightarrow \tilde {\sf D}$,
 by the normal displacement $\delta n(\xi )$
of the boundary at a boundary point $\xi$ (Fig.~\ref{fi:var}).

Consider a small variation of the potential
$W$ in the condition (\ref{large3}),
which determines the shape of ${\sf D}$
at a fixed $t$.
To take into account the deformation of the domain,
$\delta {\sf D}=\tilde {\sf D} \setminus {\sf D}$,
we write, for any fixed function $f$,
$$
\delta \left (
\oint_{\p {\sf D} }\!\! f(\zeta ) d \zeta \right ) =
\oint_{\p (\delta {\sf D})}\!\! f(\zeta ) d \zeta=2i\!\!
\int_{\delta {\sf D}}
\!\!\bar \p f(\zeta ) d^2 \zeta \approx 2i \!
\oint_{\p {\sf D}} \!\!\bar \p f(\zeta )\delta n(\zeta )
|d\zeta |
$$
and thus obtain from (\ref{large3}):
\beq\label{cor1}
\oint_{\p {\sf D}}\!\!
\frac{\p \, \delta W(\zeta ) d\zeta}{z-\zeta} \, + \,
\frac{i}{2}\oint_{\p {\sf D}}\!\!
\frac{\Delta W(\zeta ) \delta n(\zeta )}{z-\zeta}
\, |d\zeta | \, = \, 0
\eeq
Here the first term comes from the variation of $W$ and
the second one comes from the change
of ${\sf D}$.

This is an integral equation for the $\delta n (\zeta )$.
It can be solved in terms of the exterior
Dirichlet boundary value problem.
Given any smooth function $f(z)$,
let $f^H (z)$ be its
{\it harmonic continuation} (already introduced
in Section 4.2) from the boundary
of ${\sf D}$ to its exterior, i.e., the function such that
it is harmonic in ${\sf D^c}$, $ \Delta f^H =0$,
and regular at $\infty$, and
$f^H (z)=f(z)$ for all $z\in \p {\sf D}$.
The harmonic continuation is known to be unique.
Explicitly, a harmonic function
can be reconstructed from its boundary
value by means
of the {\it Dirichlet formula}
\beq\label{dirich}
f^H (z)=-\frac{1}{2\pi} \oint_{\p {\sf D}}
f(\xi )\p_n G(z, \xi) |d\xi |
\eeq
The main ingredient of this formula
is $G(z, \xi )$, which is the Green function
of the domain
${\sf D^c}$:
$$
\Delta_z G(z, \zeta )=2\pi \delta (z-\zeta )
\quad \mbox{in ${\sf D^c}$}\,,
\;\;\;\;
G(z, \zeta )=0 \quad \mbox{if $z\in \p {\sf D}$}
$$
As $\zeta \to z$, it has the logarithmic singularity
$G(z, \zeta )\to \log |z-\zeta |$. The well known
properties of harmonic functions imply that
$\p_n G(z, \xi )\leq 0$ for all $\xi \in \p {\sf D}$
(the operator of the normal derivative acts here to the
second argument).

Consider the integral
$$
\oint_{\p {\sf D}}\!\!
\frac{\p (\delta W^H) d\zeta}{z-\zeta}
$$
which is obviously equal to $0$ for all $z$ inside ${\sf D}$,
subtract it from the first term in (\ref{cor1}) and
rewrite the latter in the form
$$
\oint_{\p {\sf D}}\!\!
\frac{\p (\delta W\! - \! \delta W^H) d\zeta}{z-\zeta}
=
\frac{i}{2}\oint_{\p {\sf D}}\!\!
\frac{\p_n (\delta W \!- \! \delta W^H )}{z-\zeta}
\, |d\zeta |
$$
Here the integral over $d\zeta$ is transformed to the
integral over
the line element $|d\zeta |$. The normal derivative
is taken in the exterior of the boundary.
After this simple transformation
our condition acquires the form
\beq\label{cor2}
\oint_{\p {\sf D}}
\frac{\delta  n_1 (\zeta )+\hat R(\zeta )}{z-\zeta}
\, |d\zeta | =0 \quad \quad
\mbox{for all $z\in {\sf D}$}
\eeq
where $\delta  n_1 (z):=\Delta W(z)\delta n(z)$
(just for brevity) and $\hat R$ is the Neumann jump
operator. Acting on a smooth function $f$, this operator
gives the difference between the
normal derivative of this
function and the normal derivative
of its harmonic extension:
\beq\label{Neumann}
\hat R f (z) =\p_{n}^{-}(f(z)-f^H(z))
\eeq
The superscript indicates that the derivative is
taken in the exterior of the boundary.

By properties of Cauchy integrals, it follows from
(\ref{cor2}) that
$\Bigl [ \delta n_1 (z) + \hat R \delta W(z) \Bigr ]
\frac{|dz|}{dz}$
is the boundary value of an analytic function $h(z)$ in ${\sf D^c}$
such that $h(\infty )=0$.
For $z\in {\sf D^c}$, this function is given by
$$
h(z)=\frac{1}{2\pi i}
\oint_{\p {\sf D}}
\frac{\delta  n_1 (\zeta )+\hat R(\zeta )}{z-\zeta}
\, |d\zeta |
$$
If $h$ is not identically zero, the number of
zeros of $g$ in ${\sf D^c}$, counted with
multiplicities, is given by the
contour integral $\frac{1}{2\pi}
\oint_{\p {\sf D}}d(\mbox{arg} \, h)$.
Since $\mbox{arg} \, h (z) =\mbox{arg} ( |dz|/dz ) =-\theta (z)$,
where $\theta$ is the angle between the tangent vector
and the real axis, this integral is equal to $1$.
We conclude that the function
$h$ has exactly one simple zero outside
the domain ${\sf D}$. It is just the zero at $\infty$.

On the other hand, the variation of
the normalization condition yields,
in a similar manner:
\beq\label{cor3}
\oint_{\p {\sf D}}(\delta n_1 +\p_n \delta W )|d\zeta |=0
\eeq
This relation implies that
the zero at $\infty$ is at least of the 2-nd order. Indeed,
expanding the Cauchy integral around $\infty$,
$$
2\pi i h(z) \to \frac{1}{z}
\oint_{\p {\sf D}}(
\delta n_1 \! +\! \p_n \delta W \!- \!\p_n \delta W^H )|d\zeta |
 +O(z^{-2})
$$
one concludes,
using the Gauss law $\oint_{\p {\sf D}}
\p_n \delta W^H |d\zeta |=-\! \! \int_{{\sf D^c}}
\Delta \delta W^H d^2 \zeta =0$,
 that the coefficient in front of $1/z$
vanishes.
We have got a contradiction.

Therefore, $h(z)\equiv 0$, and
so $\delta n_1 (z)+\hat R\delta W(z)=0$. This gives
the following result for the normal
displacement of the boundary
caused by small changes of the potential
$W \rightarrow W+\delta W$:
\beq\label{deltan}
\delta n (z)=\frac{\p_{n}^{-} (\delta W^H (z)\! - \!
\delta W(z))}{\Delta W(z)}
\eeq

\paragraph{Some results for the correlation functions.}
In order to
find the correlation functions
of traces, we use the general variational formulas (\ref{var}),
where the exact free energy is replaced by the leading
contribution (\ref{free1}):
$$
\lim_{\hbar \to 0}
\lbracket \rho (z)\rbracket =
\frac{\delta F_0}{\delta W(z)} =\rhocl (z)\,,
\;\;\;\;\;
\lim_{\hbar \to 0}
\lbracket \rho (z_1)\rho(z_2)\rbracket _{c}=
\hbar^2 \, \frac{\delta \rhocl (z_1)}{\delta
W(z_2)}
$$
Basically, these are linear response relations used
in the Coulomb gas theory \cite{Forrester}.
In this approximation,
the eigenvalue plasma is represented as a continuous charged
fluid, so the information about its discrete microscopic
structure is lost.
So, these formulas
give ``smoothed" correlation functions
in the first non-vanishing order in $\hbar$. They are correct
at distances much larger than the mean distance
between the charges.

Here are the main results for the
correlation functions obtained by the variational
technique. For details of the derivation
see \cite{WZnormal} and Appendix C.

The leading contribution to the one-trace function
was already found in Section 4.2.
Here we present the general result including the first subleading
correction which can be found by variation of
(\ref{F1/2}):
\beq\label{1trace}
\begin{array}{c}
\beta
\lbracket \mbox{tr} f(\Phi ) \rbracket \,\,=\,\,
\displaystyle{
\frac{1}{\hbar}\int_{{\sf D}}
\sigma (z) f(z) \, d^2 z}
\\ \\
+\displaystyle{\frac{2\! -\! \beta}{8\pi}\left [
\int_{{\sf D}} (1\! +\! \log \sigma (z) )\Delta f(z) \, d^2 z -
\oint_{\p {\sf D}} \! \! \log \sigma (z)
\, \hat R f(z) |dz|\right ]\,
+ \, O(\hbar )}
\end{array}
\eeq
where $\hat R$ is the Neumann jump operator (\ref{Neumann}).
Applying this formula to the function $\varphi (z)$,
we get the familiar
result $\lbracket \varphi (z) \rbracket =\vphcl (z)+
\varphi_{\hbar}(z) +O(\hbar^2 )$, where $\varphi_{\hbar}$
is given by (\ref{phihbar}).
The connected two-trace function is:
\beq\label{2trace}
\beta \lbracket \mbox{tr}f \, \mbox{tr} g\rbracket _{c}
=\frac{1}{4\pi} \int_{{\sf D}} \nabla f \nabla g d^2 z-
\frac{1}{4\pi} \oint_{\p {\sf D}} f \p_n g^H |dz|
+O(\hbar )
\eeq
In particular, for the connected
correlation functions of the fields
$\varphi (z_1)$, $\varphi (z_2)$ (see (\ref{functions1}))
this formula gives
(if $z_{1,2}\in {\sf D^c}$):
\beq\label{2trace1}
\frac{1}{2\beta \hbar^2}\lbracket \varphi (z_1 )\varphi (z_2)
\rbracket_{c} =G(z_1 , z_2 ) -
G(z_1 , \infty )-G(\infty , z_2 )-
\log \frac{|z_1 - z_2 |}{r} +O(\hbar )
\eeq
where $G$ is the Green function of the Dirichlet
boundary value problem and
\beq\label{confrad}
\displaystyle{r=\exp \Bigl [ \lim_{\xi \to \infty}(
\log |\xi | +G(\xi , \infty ))\Bigr ]}
\eeq
is the (external) conformal radius of the domain ${\sf D}$.
The 2-trace functions are {\it universal}, i.e.,
they depend on the shape of the support of eigenvalues only and
do not depend on the potential $W$ explicitly.
They resemble the two-point functions
of the Hermitian 2-matrix model found in \cite{DKK}; they were
also obtained in \cite{AlJan} in the study of thermal fluctuations of a
confined 2D Coulomb gas.
The structure of the formulas indicates
that there are local correlations in the
bulk as well as strong long range correlations at the edge
of the support of eigenvalues.
(See \cite{Jancovici82} for a similar result in the context
of classical Coulomb systems).

From the mathematical point of view, the significance
of formula (\ref{2trace1}) is to provide a link between
such seemingly unrelated disciplines as
classical analysis in two dimensions and the random matrix
theory. Namely, different limits or certain specifications of the
arguments in this formula allow one
to represent some important objects of classical analysis
associated with the domains ${\sf D}$
and ${\sf D^c}$ (e.g., the conformal map
onto the unit circle and its Schwarzian derivative,
the Bergman kernel) in terms of correlations between
eigenvalues of random matrices.

Further variation of the pair density correlation
function suggests that, starting from $n =3$, the connected
$n$-point density correlations
vanish in the bulk in all orders of $\hbar$
(in fact they are exponential in $1/\hbar$).
The entire leading contribution
comes from the boundary.
The result for the connected three-trace function is:
\beq\label{3trace}
\beta \lbracket \prod_{i=1}^{3}\mbox{tr}\, f_i \rbracket_c
=\frac{\hbar}{16\pi^2}\oint_{\p {\sf D}}
\frac{|dz|}{\sigma (z)}
\prod_{j=1}^{3}\hat Rf_j (z) +O(\hbar^2 )
\eeq

\section{The matrix model as a growth problem}

\subsection{Growth of the support of eigenvalues}

When $N$ increases at a fixed potential $W$,
one may say that the support of eigenvalues grows.
More precisely, we are going to find how
the shape of the support of eigenvalues
changes under
$t \rightarrow t+\delta t$, where $t=N\hbar$,
if $W$ stays fixed.

The starting point is the same as for the variations
of the potential, and the
calculations are very similar as well.
Variation of the condition (\ref{large3})
and of the normalization
condition yields
$$
\oint_{\p {\sf D}}\frac{\Delta W(\zeta )\delta n(\zeta )}{z-\zeta}\,
|d\zeta| =0 \;\;\;\mbox{(for all $z\in {\sf D}$)},
\;\;\;\;\;
\oint_{\p {\sf D}}\Delta W(\zeta )\delta n(\zeta ) \, |d\zeta |
=-4\pi \beta \delta t
$$
The first equation means that
$\Delta W(z)\delta n (z)
\frac{|dz|}{dz}$
is the boundary value of an analytic function
$h(z)$ such that
$h(z)=-4\pi \beta \delta t/z +O(z^{-2})$
as $z \to \infty$.
The solution for the $\delta n (z)$ is:
\beq\label{gr0}
\delta n(z)=-\, \frac{\beta \delta t}{2\pi \sigma (z)} \,
\p_n G(\infty , z)
\eeq
where $G$ is the Green function of the
Dirichlet boundary problem
in ${\sf D^c}$.
For quasiharmonic potentials (with $\sigma =1/\pi$),
the formula simplifies:
\beq\label{gr1}
\delta n(z)=-\frac{\beta}{2} \, \delta t\,
\p_n G(\infty , z)
\eeq
Identifying $t$ with time,
one can say
that the normal velocity of the boundary,
$V_n = \delta n / \delta t$, is
proportional to gradient of the Green function:
$V_n \propto -\p_n G(\infty , z)$.
This result is quite general. It holds for any
(not necessarily connected) domains of eigenvalues with
a smooth boundary.

If the domain is connected,
the Green function can be expressed through the conformal
map $w(z)$ from ${\sf D^c}$ onto the exterior
of the unit circle:
\beq\label{Gw}
G(z_1 , z_2 )=\log \left |
\frac{w(z_1 )-w(z_2 )}{1-w(z_1 ) \overline{w(z_2 )}}
\right |
\eeq
In particular,
$G(\infty , z) =-\log |w(z)|$.
As $|z|\to \infty$,
$w(z)=z/r + O(1)$, where $r$ is the external conformal radius
of the domain ${\sf D}$ which enters eq. (\ref{2trace1}).
It is easy to see that $\p_n \log |w(z)| = |w'(z)|$ on $\p {\sf D}$,
so one can rewrite the growth law (\ref{gr1}) as follows:
\beq\label{gr1a}
\delta n(z)=\frac{\beta}{2}\delta t \,
|w'(z)|
\eeq

\begin{figure}[tb]
\epsfysize=6cm
\centerline{\epsfbox{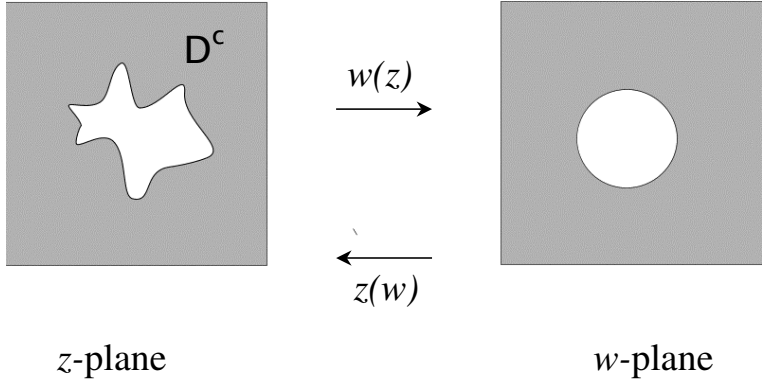}}
\caption{\sl The mutually inverse conformal maps
$w(z)$ and $z(w)$.}
\label{fi:maps}
\end{figure}

It is worth noting that
the inverse map, $z(w)$, is the classical ($\hbar \to 0$)
limit of the Lax operator (\ref{Lax}) of the 2D Toda
hierarchy.
Indeed, making the rescaling
$n\hbar =t$, $\p /\p n =\hbar \p_t$, we see that the shift
operator
$e^{\hbar \p_t}$ can be replaced by a commuting variable $w$
with the Poisson bracket
$\{\log w ,\, t\}=1$. In this limit, known also as the
dispersionless limit of the 2D Toda hierarchy,
the Lax operator $L(e^{\hbar \p_t})$ converts into
the function $z(w)$ given by the series of the form
\beq\label{z(w)}
z(w)=r(t) w +\sum_{k\geq 0} u_k (t) w^{-k}
\eeq
It defines the one-to-one conformal map from the exterior
of the unit circle onto ${\sf D^c}$, which is inverse to
the map $w(z)$ (Fig.~\ref{fi:maps}).
For more details see \cite{teodor}.

The basic formula which allows one to find the $t$-derivative
of any quantity of the form $\int_{{\sf D}(t)} f(z) \, d^2 z$
(where the function $f$ is assumed to be independent of $t$)
immediately follows from (\ref{gr0}):
$$
\delta \left ( \int_{{\sf D}} f(z) \, d^2 z \right )
=\oint_{{\sf D}} f(z) \delta n(z) \, |dz|=
-\, \frac{\beta \delta t}{2\pi}
\oint_{{\sf \p D}} \frac{f(z)}{\sigma (z)}\p_n G(\infty , z)
 \, |dz|
$$
In the r.h.s. we recognize the value at infinity of
the harmonic continuation of the function $f(z)/\sigma (z)$,
so the result is
\beq\label{gr2}
\frac{\p}{\p t}
\left ( \int_{{\sf D}} f(z) \, d^2 z \right )
=\beta \left ( f/\sigma \right )^H (\infty )
\eeq
Using this formula and the integral representation
of $F_0$, we find $t$-derivatives of the free energy.
The first derivative,
\beq\label{gr3}
\p_t F_0 = 2 \int_{{\sf D}} \sigma (z) \log |z|\,
d^2 z
\eeq
was already found in Section 4.3
by other means (see (\ref{gr3a})).
The second derivative is proportional
to the logarithm of the conformal radius
(\ref{confrad}):
\beq\label{gr4}
\left. \phantom{\frac{a}{b}}
\p_{t}^{2} F_0 = 2 \beta
\left ( \log |z| \right )^H \right|_{z=\infty}
=2\beta
\lim_{z\to \infty}\left ( \log |z| +G(z, \infty )\right )
=2\beta \log r
\eeq

In this connection
let us also mention
the nice formula for the conformal map
$w(z)$,
\beq\label{limpsi}
w(z)=\lim_{N \to \infty} \, \frac{\psi_{N+1}(z)}{\psi_{N}(z)}
\eeq
which follows from the relation
$\log \left (P_{N+1}(z)/P_{N}(z)\right )=
\hbar \p_t \lbracket \mbox{tr} \log (z-\Phi )\rbracket
+O(\hbar )$ after calculating the $t$-derivative according to
the above rule. This connection between
conformal maps and orthogonal polynomials
goes back to the classical theory of analytic functions
(see e.g. \cite{conformal}). It is the
context of the random matrix theory where
it looks natural and easily understandable.

\subsection{Laplacian growth}

The growth law (\ref{gr1}) is common to many important problems
in physics.
The class of growth processes, in which
dynamics of a moving front (an interface) between two distinct phases
is driven by a harmonic scalar field
is known under the name {\it Laplacian growth}.
The most known examples are
viscous flows in the Hele-Shaw cell (the Saffman-Taylor problem),
filtration processes in porous media,
electrodeposition and
solidification of undercooled liquids.
A comprehensive list of relevant papers
published prior to 1998  can be found in
\cite{list}.
Recently, the Laplacian growth mechanism was
recognized \cite{Agametal} in a purely quantum
evolution of semiclassical electronic blobs
in the QH regime.

\paragraph{The Saffman-Taylor problem.}
Let us describe the main features of the Laplacian growth
on the example of viscous flows.
To be specific, we shall speak about an interface between two
incompressible fluids with very different viscosities on the plane
(say, oil and water).
In practice, the 2D geometry is
realized in the Hele-Shaw cell -- a narrow gap between two parallel
glass plates.
In this version, the problem is also  known as the Saffman-Taylor problem
or viscous fingering. For a review, see \cite{RMP}.
The velocity field in a viscous fluid in the Hele-Shaw cell
is proportional to the gradient of pressure $p$ (Darcy's law):
$$
\vec V =-K \nabla p\,,
\;\;\;\;
K=\frac{b^2}{12 \mu}
$$
Here the constant
$K$ is called the filtration coefficient, $\mu$ is viscosity and
$b$ is the size of the gap between the two plates.
Note that if $\mu \to 0$, then $\nabla p \to 0$, i.e.,
pressure in a fluid with negligibly small viscosity is
uniform. Incompressibility of the fluids ($\nabla \vec V =0$)
implies that the
pressure field is harmonic: $\Delta p =0$.
By continuity, the velocity of the interface
between the two fluids is proportional to the normal
derivative of the pressure field on the boundary:
$V_n =-K \p_n p$.

\begin{figure}[tb]
\epsfysize=6cm
\centerline{\epsfbox{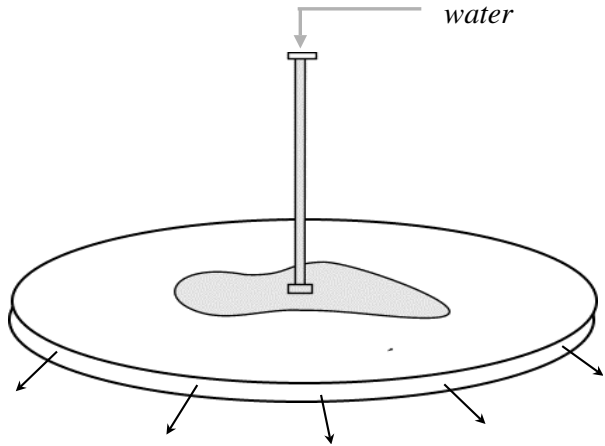}}
\caption{\sl
The Hele-Shaw cell.}
\label{fi:Hele-Shaw}
\end{figure}

To be definite, we assume that the Hele-Shaw cell
contains a bounded droplet
of water surrounded by an infinite ``sea" of oil
(another possible
experimental set-up is an air bubble surrounded
by water).
Water is injected into the droplet while
oil is withdrawn at infinity
at a constant rate, as is shown
schematically in Fig.~\ref{fi:Hele-Shaw}.
The latter means that the pressure
field behaves as $p \propto -\log |z|$ at large distances.
We also assume that the interface between oil and water
is a smooth closed curve $\gamma$ which depends on time.
As it was mentioned above,
if viscosity of water is negligible, then one may set
$p=0$ inside the water droplet. However, pressure usually
has a jump across the interface, so $p$ in general
does not tend to zero if one approaches
the boundary from outside. This effect is due to
{\it surface tension}. It is hard to give realistic
estimates of the surface tension effect from first principles, so
one often employs certain ad hoc assumptions.
The most popular one is to say that
the pressure jump is proportional to the local curvature
of the interface.

To summarize, the
mathematical setting of the Saffman-Taylor problem
is as follows:
\beq\label{lg1}
\left \{
\begin{array}{ll}
V_n =-\p_n p & \quad \mbox{on $\gamma$}
\\
\Delta p =0 &\quad \mbox{in oil}
\\
p\to -\log |z| & \quad \mbox{in oil as $z\to \infty$}
\\
 p=0 & \quad \mbox{in water}
\\
p^{(+)}-p^{(-)}=-\nu \kappa & \quad \mbox{across $\gamma$}
\end{array}
\right.
\eeq
Here $\nu$ is
the surface tension coefficient and
$\kappa$ is the local curvature of the interface.
(The filtration coefficient is set to be $1$.)
The experimental evidence suggests that
when surface tension is small enough,
the dynamics becomes unstable. Any initial
domain develops an unstable fingering pattern.
The fingers split into new ones, and after a
long lapse of time the water droplet
attains a fractal-like structure.
This phenomenon is similar to
the formation of fractal patterns
in the diffusion-limited aggregation.

Comparing (\ref{gr1}) and (\ref{lg1}),
we identify the ${\sf D}$ and ${\sf D^c}$ with
the domains occupied by water and oil
respectively, and conclude that the growth laws are
identical, with the pressure field being
given by the Green function:
$p(z) =G(\infty , z)$, and $p=0$ on the interface.
The latter means that supports of eigenvalues
grow according to (\ref{lg1})
with {\it zero surface tension}, i.e.,
with $\nu =0$ in (\ref{lg1}).

Neglecting the surface tension effects, one
obtains a good approximation unless the curvature
of the interface becomes large.
We see that the idealized
Laplacian growth problem, i.e., the one
with zero surface tension, is mathematically equivalent
to the growth of the support of eigenvalues
in ensembles of random matrices $\normal$, $\normal^0$
and $\complex$. This fact clarifies the origin of
the integrable structure of the Laplacian growth with zero surface
tension discovered in \cite{M-WWZ}.
The link to the normal
matrix model has been established in \cite{KKMWZ},
see also \cite{Zabmatrix}.

\paragraph{The finite-time singularities.}
As a matter of fact,
the Laplacian growth problem with zero surface tension
is ill-posed since an initially smooth interface
often becomes singular in the process of evolution,
and the solution blows up.
The role of surface tension is to inhibit
a limitless increase of the interface curvature.
In the absense of such a cutoff, the tip of
the most rapidly growing finger typically grows
to a singularity (a cusp).
In particular, a singularity necessarily occurs
for any initial interface that is the image of the
unit circle under a rational conformal map,
with the only exception of an ellipse.

An important fact is
that the cusp-like singularity occurs at a finite time $t=t_c$,
i.e., at a finite area of the droplet.
It can be shown that the conformal radius of the droplet $r$
(as well as some other geometric parameters),
as $t\to t_c$, exhibits a singular behaviour
$$
r-r_c \propto (t_c -t)^{-\gamma}
$$
characterized by a critical exponent $\gamma$.
The generic singularity is the cusp $(2,3)$,
which in suitable local coordinates looks like
$y^2 =x^3$. In this case $\gamma =-\frac{1}{2}$.
The evolution can not be extended beyond $t_c$.

A similar phenomenon was well-known in the theory
of random matrices for quite a long time, and in fact it
was the key to their applications to 2D quantum gravity and string
theory. In the large $N$ limit, the random matrix models
have {\it critical points} -- the points
where the free energy is not analytic as a function
of a coupling constant. As we have
seen, the Laplacian growth time $t$
should be identified with a coupling constant
of the normal or complex matrix model.
In a vicinity of a critical point,
$$
F_0 \sim F_{0}^{{\rm reg}} \, + \alpha  (t_{c}-t)^{2-\gamma}
$$
where the critical index $\gamma$ (often
denoted by $\gamma_{{\rm str}}$ in applications to string theory)
depends on the type of the critical point.
Accordingly, the singularities show up in correlation
functions. Using the equivalence established above,
we can say that
the finite-time blow-up (a cusp-like singularity)
of the Laplacian growth with zero surface tension
is a {\it critical point} of the
normal and complex matrix models.

\subsection{The semiclassical limit for electrons in
magnetic field}

As we have seen in Section 2.3,
the system of $N$ electrons in the plane in non-uniform
magnetic field, which fully occupy the lowest energy level,
is equivalent to the ensemble of normal $N\times N$ matrices,
where $N$ is degeneracy of the level.
In this section we study
the QH droplet in the semiclassical
regime. The equivalence with the large
$N$ limit of the matrix model suggests to identify
the semiclassical QH droplet
with the support of eigenvalues.
Remarkably, it is this limit where one makes contact with
the purely classical Saffman-Taylor problem.

A remark is in order. The limit $\hbar \to 0$ we
are talking about is really a {\it semi}classical
limit, or better to say ``partially classical".
Although the Planck constant $\hbar$ tends to zero,
all the particles remain at the
lowest (most quantum) energy level,
assuming their mass is small or the magnetic field
is large. The true quasiclassical limit would
imply that the particles
occupy higher energy levels.

We recall that the joint probability to find electrons
at the points $z_i$ is $|\Psi_N (z_1 , \ldots , z_N)|^2$,
with $\Psi_N \propto \det_{N\times N} [\psi_{n}(z_k)]$,
where
\beq\label{semi1}
\psi_n (z)=\frac{1}{\sqrt{h_{n-1}}}P_{n-1}
(z) e^{W(z)/(2\hbar)}
\eeq
are orthogonal one-particle wave functions
for electrons in the magnetic field
$B=-\frac{1}{2}\Delta W$ at the lowest level
$E=0$.
The level is assumed to be completely filled, i.e.,
$n=0, 1 \ldots , N=[\phi /\phi_0 ]$, where
$\phi_0$ is the flux quantum.
Then the mean density of the electrons coincides with
the expectation value of the density of eigenvalues
in the normal or complex matrix model.

\begin{figure}[tb]
\epsfysize=6.5cm
\centerline{\epsfbox{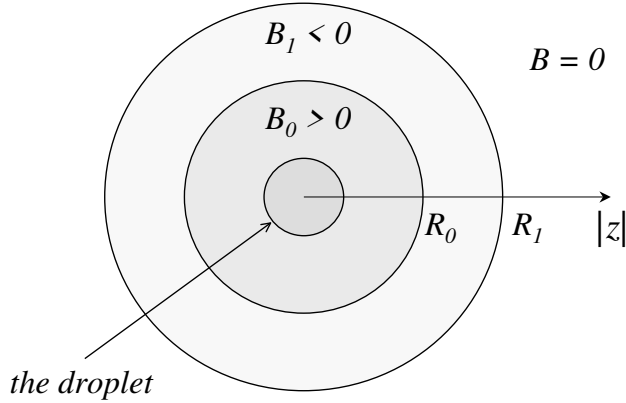}}
\caption{\sl The configuration of magnetic fields. }
\label{fi:arrangement}
\end{figure}

The degeneracy of the level cam be controlled
in different ways. One of them is to assume the
following arrangement
(see Fig.~\ref{fi:arrangement}).
Let a strong uniform magnetic field
$B_0 >0$ be applied in a large disk of radius $R_0$.
The disk is surrounded by a large annulus $R_0 < |z|< R_1$
with a magnetic field $B_1 <0$ such that the total magnetic flux
through the system is $N\phi_0$:
$\pi B_0 R_{0}^{2}
-\pi B_1 (R_{1}^{2} -R_{0}^{2})=N \phi_0$.
The magnetic field outside the largest disk $|z|<R_1$
vanishes. The disk is connected through a tunnel barrier
to a large capacitor that maintains
a small positive chemical potential
slightly above the zero energy.
If the field $B_0$ is strong enough,
the gap between the energy levels is large,
and the higher levels can be neglected.
In the case of the uniform fields
$B_0$ and $B_1$ the QH droplet is a disk
of radius $r_0 =\sqrt{N\hbar}\ll R_0$
trapped at the origin.

In this set-up, let us apply a non-uniform magnetic field,
$\delta B$, somewhere inside the disk $|z|<R_0$ but
well away from the droplet.
This leads to the following two effects.

\paragraph{The Aharonov-Bohm effect in the QH regime
\cite{Agametal,Wiegmann}.}
Suppose that the nonuniform magnetic field $\delta B$
does not change the total flux: $\int \delta B d^2z =0$.
As is argued above,
the shape of the droplet
is the same as that of the support of eigenvalues in the
ensemble $\normal$ with the potential
$$
W(z)=-\frac{B_0}{2}|z|^2 -\frac{1}{\pi}
\int \log |z-\zeta |\, \delta B (\zeta )\, d^2 \zeta
$$
The second term is harmonic inside and around the droplet.
One may have in mind thin solenoids
carrying magnetic flux (``magnetic impurities").
In the case of point-like magnetic fluxes
$q_i$ at points $a_i$ the change of the potential is
$\delta W(z)=
\sum_i q_i \log |z-a_i |$.

\begin{figure}[tb]
\epsfysize=4cm
\centerline{\epsfbox{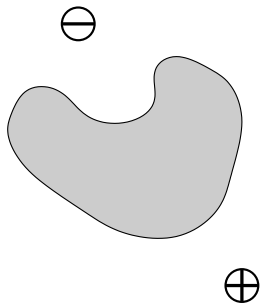}} \caption{\sl The
electronic droplet in the presence of magnetic impurities. }
\label{fi:impur}
\end{figure}

Let us stress that in the presence of the fluxes,
the shape of the droplet is no longer circular
although the magnetic field inside the droplet
and not far from it
remains uniform and is not changed at all
(Fig.~\ref{fi:impur}). In this respect
this phenomenon is similar to the Aharonov-Bohm
effect. Due to the quantum
interference the electronic fluid is attracted
to positive fluxes and is repelled
by negative ones.
The response of the droplet to an
infinitesimal change of the magnetic field
$\delta B$ is described by eq.\,(\ref{deltan})
in which
$$
\delta W^H(z) -\delta W(z)=
\frac{1}{\pi}\int_{{\sf D^c}}
G(z, \zeta )\delta B(\zeta ) d^2 \zeta
$$
In fact this formula holds for arbitrary
$\delta B$, not necessarily vanishing inside the droplet.
In particular, for small point-like fluxes
$\delta q_i$ at some points $a_i$ we have
$\delta W =\sum_i \delta q_i \log |z-a_i|$,
$\delta B=-\pi \sum_i \delta q_i \delta^{(2)}(z-a_i)$, and
$\delta W^H(z) -\delta W(z)=
-\sum_i  \, G(z, a_i)\delta q_i$.
If $a_i$ is inside, $G(z, a_i)$ is set to be zero.
The sum, therefore, is over outside fluxes only.
The fluxes inside the droplet, if any,
appear to be completely screened and do not have any influence
on its shape.

\paragraph{Growth of the electronic droplet.}
If the total magnetic
flux increases,
with magnetic impurities kept fixed, the electronic
droplet grows. For example, one may adiabatically increase
$B_1$, with
$B_0$ and $\delta B$ fixed. Then the droplet grows because
the degeneracy of the lowest level is enlarged and new
electrons enter the system.
The growth is described by eq.\,(\ref{gr0})
with $\Delta W (z)=- 2 B_0$ which is
equivalent to the Darcy law.
This phenomenon is purely quantum.
Like the Aharonov-Bohm effect, it is caused by
quantum interference. Its characteristic scale
is less than that of the Saffman-Taylor fingering
by a factor of $10^9$.
The correspondence established above suggests that the edge of the
QH droplet may develop unstable features similar to the
fingers in the Hele-Shaw cell.

\subsection{Semiclassical $\psi$-function}

In this Subsection we derive
the semiclassical asymptotics of the $\psi$-function
(\ref{semi1}), i.e.,
$$\psi_{N+1}(z)\stackrel{\hbar \to 0}{\longrightarrow}
\psi (z)
$$
To avoid cumbersome
technical details, we mainly consider models with
quasiharmonic potential.

First of all, it is necessary
to know the large $N$
limit of the orthogonal polynomials $P_N$.
Truncating the general formula (\ref{expf})
at the second term in the exponent, we can write:
\beq\label{psi1}
\begin{array}{c}
\displaystyle{P_N (z)= \lbracket \det (z-\Phi )\rbracket
=\lbracket e^{\mbox{tr}\, \log (z-\Phi )}\rbracket }
\\ \\
\displaystyle{
\stackrel{\hbar \to 0}{=}\,\,
\exp \left (\lbracket  \mbox{tr}\, \log
 (z-\Phi )\rbracket  +\frac{1}{2}
\lbracket  \left (\mbox{tr}\, \log
 (z-\Phi )\right )^2 \rbracket_c +\ldots \right )}
\end{array}
\eeq
The first term in the r.h.s. is $O(\hbar^{-1})$, the second one is
$O(\hbar^0)$ and the ignored terms vanish as $\hbar \to 0$.
However, this formula
should be applied with some care since the logarithm is
not a single-valued function.
This formula is correct if one can
fix a single-valued
branch of the logarithm.
It is possible if $z$ is outside ${\sf D}$.
For $z$ inside ${\sf D}$ an
analytic continuation should be used.
To take care of the normalization, we
also need
\beq\label{hN}
h_N =  \frac{Z_{N+1}}{(N\! +\! 1)Z_N} =
(2\pi^3 \hbar )^{1/2}
\exp \left ( \hbar^{-1} \p_t F_0 +\frac{1}{2}
\p_{t}^{2}F_0  + \ldots \right )
\eeq
which is written here with the same precision
as (\ref{psi1}).
(We have used (\ref{hbarexp1}) with
$\alpha =\log\sqrt{2\pi^3 \hbar}$ and
have taken into account that $F_{1/2}=0$
for quasiharmonic potentials.)

Let us first keep the dominant terms in the r.h.s. of
(\ref{psi1}), (\ref{hN}) and ignore the
$O(\hbar^0)$ terms for a while.
We have, for $z \in {\sf D^c}$:
$|\psi (z)|^2 \sim
e^{-\frac{1}{\hbar} \p_t F_0 }
\, e^{\frac{1}{\hbar}(W(z)-\lbracket \varphi (z)
\rbracket )}$,
or, using the results of the previous section,
$$
|\psi (z)|^2 \sim
e^{-2{\cal A}(z)/\hbar}\,,
\;\;\;\;\;
2{\cal A}(z)=\vphcl (z) -\vphcl (0)-W(z)
$$
As we know, ${\cal A}(z)$ defined by this formula
is zero on the boundary.
The analytic continuation of the ${\cal A}(z)$ inside
the contour $\gamma =\p {\sf D}$
can be done using the Schwarz function.
Namely, since $2 \p {\cal A}(z) = \bar z - S(z)$,
the desired analytic continuation can be defined
(up to a constant) by the formula
$$
2 {\cal A}(z)=|z|^2 -2{\cal R}e  \int^z S(\zeta )\, d\zeta
$$
Clearly, ${\cal A}(z)$ defined in this way
is constant on the contour $\gamma$.
Indeed,
if $z_{1,2}\in \gamma$, then
$$
2({\cal A}(z_2 )-{\cal A}(z_1 ))=
|z_2|^2 -|z_1 |^2 -2\Re \int_{z_1}^{z_2} S(z)dz
$$
$$
=\, \, z_2 \bar z_2 -z_1 \bar z_1 -\int_{z_1}^{z_2}\bar z dz
-\int_{z_1}^{z_2} z d\bar z
=\, \, z_2 \bar z_2 -z_1 \bar z_1 -
\int_{z_1}^{z_2}d(z \bar z)\, =\, 0
$$
Since ${\cal A}(z)$ should be zero on $\gamma$,
we finally define
\beq\label{A(z)}
{\cal A}(z)=\frac{1}{2}|z|^2 -
\frac{1}{2}|\xi_0 |^2 -
\Re \int^{z}_{\xi_0} S(\zeta )d\zeta
\eeq
where $\xi_0$ is an arbitrary point on the contour.
We call the function defined by (\ref{A(z)}) {\it
the effective action}.
From the above it follows that its
first derivatives vanish for all $z\in \gamma$:
$\p {\cal A}(z)=\bar \p {\cal A}(z)=0$.
This means that $|\psi |^2 \sim
e^{-2{\cal A}/\hbar}$ has a sharp
maximum on the contour $\gamma$.
We may say that purely quantum particles are in general delocalized
in the plane,
purely classical particles are localized at some points
in the plane
while partially classical particles, like our
electrons, are localized on closed curves in the plane.

Let us turn to the $O(\hbar^0 )$-corrections, which
give the subexponential factor
in the asymptotics of the $\psi$-function\footnote{In what
follows we
restrict our consideration
to the squared modulus $|\psi (z)|^2$. The
$\psi$-function itself contains a phase factor which
rapidly oscillates in both tangential and normal directions.
(I thank A.Bo\-yar\-sky and
O.Ru\-chai\-sky for a discussion on this point.)}.
We assume, for simplicity, that the domain ${\sf D}$ is connected.
Extracting analytic and anti-analytic parts of
eq. (\ref{2trace1}), we get
$$
\lbracket (\mbox{tr}\,
\log (z-\Phi ))^2 \rbracket_c =
\log (r w'(z)) +O(\hbar )
$$
where $r$ is the external conformal radius
of ${\sf D}$ and $w'(z)$ is the
derivative of the conformal map from ${\sf D^c}$
onto the exterior of the unit circle. Plugging this into
(\ref{psi1})
and taking into account eq. (\ref{gr4}), we
finally obtain:
\beq\label{psi2}
|\psi (z)|^2 = \frac{|w'(z)|}{\sqrt{2\pi^3 \hbar}}
\, e^{-2{\cal A}(z)/\hbar}
\eeq
This formula does resemble the WKB asymptotics in quantum
mechanics. If a singularity of the conformal map is
sufficiently close to the boundary from inside,
the asymptotics becomes invalid in this region.

The effective action
can be expanded near the contour, where it takes
the minimal value:
\beq\label{expanA}
{\cal A}(z+\delta_n z)=|\delta_n z |^2 \mp \frac{1}{3}\kappa (z)
|\delta_n z|^3 + \frac{1}{4} \kappa^2 (z) |\delta_n z |^4 +
\ldots
\eeq
Here $\kappa (z)$ is the local curvature of the contour
at the point $z$ and $\delta_n z$ is a small deviation
from the point $z \in \gamma$ in the normal direction.
(The upper and lower signs correspond to the
outward and inward deviations respectively.)
A similar expansion of $\log |w'(z)|$ reads
\beq\label{expw}
\log |w'(z +\delta_n z)|=\log |w'(z)| \pm
(|w'(z)| -\kappa (z))\, |\delta_n z| \, + \, \ldots
\eeq
Some details of the derivation are given in
Appendix E.
Therefore, if $\kappa (z) \ll \hbar^{-1/2}$,
the squared modulus of the $\psi$-function is well
approximated by the sharp Gaussian distribution in the
normal direction with the amplitude slowly
modulated along the curve:
\beq\label{psi3}
|\psi (z+\delta_n z)|^2 \simeq
\frac{|w'(z)|}{\sqrt{2\pi^3 \hbar}}\,
e^{-2|\delta_n z|^2 /\hbar}
\eeq
In the case of general potentials the calculations
lead to a similar result:
\beq\label{psi4}
|\psi (z+\delta_n z)|^2 \simeq
\sqrt{\frac{\sigma (z)}{2\pi^2 \hbar}}
\, |w'(z)|\,
e^{-2\pi \sigma (z) |\delta_n z|^2 /\hbar}
\eeq
We see that the width of the Gaussian distribution
depends on the point of the curve through the function
$\sigma$ (which is proportional to the magnetic field
in the QH interpretation).
Note that this asymptotics is consistent
with the normalization
$\int |\psi (z)|^2 d^2z =1$.
The easiest way to see this is to notice that
the Gaussian function in (\ref{psi3}) (as well as
the one in (\ref{psi4})) tends to the delta function
$\delta (z; \gamma )$ with the support on the curve
(see Appendix B). Hence
one can formally write the limiting $\psi$-function
in the form
\beq\label{psi5}
|\psi (z)|^2 = \frac{|w'(z)|}{2\pi} \, \delta (z; \gamma )
\eeq
and the right normalization is transparent.

Finally, we note that the growth law (\ref{gr1a})
(the Darcy law) can be written in the suggestive form
\beq\label{psi6}
V_n (z)\propto |\psi (z)|^2
\eeq
The normal velocity $V_n$ is defined by
the relation
$$
\p_t \Theta (z; {\sf D}(t))=
V_n \delta (z; \p {\sf D}(t))
$$
where $\Theta (z; {\sf D})$ is the characteristic function
of the domain ${\sf D}$.
Since the classical value
of $ \lbracket \rho (z)\rbracket$ is
$\Theta (z; {\sf D})/\pi $, we can represent the Darcy
law in yet another form:
$$
\p_t \lbracket \rho (z)\rbracket =
|\psi (z)|^2
$$
In fact it is the $\hbar \to 0$ limit of the exact
relation
\beq\label{q-darcy}
\lbracket \rho (z)\rbracket _{N}
-\lbracket \rho (z)\rbracket _{N-1}
=\hbar |\psi_N (z)|^2
\eeq
which immediately follows from
the fact that $\lbracket \rho (z)\rbracket _{N}=
\hbar \kernel_N (z, \bar z)$ and
from the definition of the kernel
function (\ref{kernel}).
In fact it
is equivalent to the Hirota equation (\ref{Hirota2}).
We see that (\ref{q-darcy})
can be regarded as a ``quantization"
of the Darcy law.

\section*{Acknowledgments}

These lectures were presented at the Les Houches
School ``Applications of Random Matrices in Physics"
in June 2004. It is a pleasure to acknowledge
the nice and productive atmosphere of the School.
I am grateful to
O.Agam,
E.Bet\-tel\-heim,
I.Kos\-tov,
I.Kri\-che\-ver,
A.Mar\-sha\-kov,
M.Mi\-ne\-ev-\-Wein\-stein,
R.Teo\-do\-res\-cu
and P.Wieg\-mann
for collaboration and
to
A.Aba\-nov,
A.Bo\-yar\-sky,
L.Chek\-hov,
A.Gor\-sky,
V.Ka\-za\-kov,
O.Ru\-chai\-sky
for discussions.
This work was supported in part by RFBR grant 04-01-00642,
by grant for support of scientific schools
NSh-1999.2003.2 and by grant INTAS 03-51-6346.

\section*{Appendices}

\subsection*{Appendix A}

\paragraph{The proof of the determinant
representation (\ref{det}).}
It is simpler to start with $\det C_{ij}$,
where $C_{ij}$ is given in (\ref{det1}), and
to show that $N!\det C_{ij}$ coincides with $Z_N$.
The proof is a chain
of obvious equalities. By definition,
$$
N! \det C_{ij}=
\sum_{Q,P}(-)^Q (-)^P \,
C_{Q(1), P(1)} \, \ldots \,
C_{Q(N), P(N)}
$$
Plugging the explicit form of the $C_{ij}$,
we write:
$$
N! \det C_{ij}=\sum_{Q,P}(-)^Q (-)^P \!\! \int \! \! \prod_{i=1}^{N}
\left [ z_{i}^{Q(i)-1} \bar z_{i}^{P(i)-1}
e^{W(z_i)}d^2 z_i \right ]
$$
and, interchanging the order of summation and integration,
obtain the result:
$$
N! \det C_{ij}=\int \underbrace{\left [
\sum_Q (-)^Q \prod_{i=1}^{N} z_{i}^{Q(i)-1}\right ]}_{
\Delta_N (z_i)}
\underbrace{\left [ \sum_P (-)^P \prod_{i=1}^{N} \bar z_{i}^{P(i)-1}
\right ]}_{\overline{\Delta_N(z_i)}}
\prod_{k=1}^{N} e^{W(z_k)} d^2 z_k \,=\,  Z_N
$$

\paragraph{Orthogonality of the polynomials (\ref{orthog1}).}
The definition (\ref{orthog1}) implies
$$
P_n(\lambda )=
\frac{1}{Z_n}\int |\Delta_n (z_i)|^2
\prod_{j=1}^{n}(\lambda -z_j)e^{W(z_j)}d^2 z_j
$$
We want to show that these polynomials are
orthogonal in the complex plane.

Set $e^{W(z)}d^2 z =d \mu$ for brevity.
It is enough to show that $\int P_n (z) \bar z^m d\mu =0$
for all $m<n$, i.e.,
$$
\int d\mu (z)\bar z^m \int |\Delta_n (z_i)|^2
\prod_{j=1}^{n}(z-z_j)d\mu (z_j)=0
$$
Note that
$\Delta_n (z_1, \ldots , z_n)\prod_{j=1}^{n}(z-z_j)=
\Delta_{n+1}(z_1 , \ldots , z_n , z)$.
Setting $z\equiv z_{n+1}$, we have:
$$
\mbox{LHS}\, =\int
\Delta_{n+1}(z_i)\overline{\Delta_{n}(z_i)}\,
\bar z_{n+1}^{m}
\prod_{j=1}^{n+1}d\mu (z_j)
$$
$$
=\frac{1}{n+1}\sum_{l=1}^{n+1} (-1)^{l+n+1}
\int \Delta_{n+1}(z_1, \ldots , z_{n+1})\,
\overline{\Delta_n (z_1 , \ldots ,\not {z_l},
\ldots , z_{n+1})z_{l}^{m}} \prod_{j=1}^{n+1}
d\mu (z_j)
$$
One can notice that the summation
gives the expansion
of the determinant
$$
\left |
\begin{array}{ccccc}
1& z_1 & \ldots & z_{1}^{n-1} & z_{1}^{m}
\\
1& z_2 & \ldots & z_{2}^{n-1} & z_{2}^{m}
\\
\ldots & \ldots & \ldots & \ldots & \ldots
\\
1& z_{n+1} & \ldots & z_{n+1}^{n-1} & z_{n+1}^{m}
\end{array}
\right |
$$
under the bar.
If $m<n$, it vanishes.
If $m=n$, it equals $\Delta_{n+1}(z_i)$, and we get
$\mbox{LHS}=\frac{1}{n+1} \, Z_{n+1}$.

\subsection*{Appendix B}

Here we list some standard formulas often used
in Section 4.

\paragraph{The complex notation.}
\begin{itemize}
\item
Complex coordinates: $z=x+iy$, $\bar z =x-iy$,
$\p_z =\frac{1}{2}(\p_x -i\p_y )$,
$\p_{\bar z} =\frac{1}{2}(\p_x +i\p_y )$.
\item
The Laplace operator:
$
\Delta =\p_{x}^{2}+\p_{y}^{2} =4\p_{z}\p_{\bar z}
$
\item
Contour integrals: let $f$ and $g$ be any smooth functions
defined in some neighborhood of the contour $\gamma$,
then
$$
\oint_{\gamma}g \p_n f |dz|=
-2i \oint_{\gamma}g \p_z f dz
-i\oint f dg
$$
\end{itemize}

\paragraph{Singular functions.}
\begin{itemize}
\item
Two-dimensional $\delta$-function:
$\delta (z)=\frac{1}{2\pi}\Delta \log |z| =
\frac{1}{\pi} \,\p_{\bar z} (1/z)$.
The characteristic property of the delta-function is
$f(z)\delta (z-a)\, d^2 z = f(a)$ for any (smooth)
function $f$.

\item
The $\delta$-function with the support
on a curve (a closed contour) $\gamma$: a function
$\delta (z;\gamma )$ such that
$$
\int f(z)\delta (z; \gamma )d^2z =\oint_{\gamma} f(z)|dz|
$$
for any smooth function $f$.
\item
The ``normal derivative'' of
the $\delta$-function of the
contour $\gamma$: a function
$\delta '(z;\gamma )$ such that
$$
\int f(z)\delta '(z; \gamma )d^2z =-\oint_{\gamma}\p_n f(z)|dz|
$$
for any smooth function $f$, with the normal vector
being directed to the exterior of the contour.
\item
The characteristic function of the domain ${\sf D}$:
$\Theta (z; {\sf D})=1$ if $z\in {\sf D}$ and $0$
otherwise;
$\nabla \Theta (z;{\sf D})=-
\vec n \, \delta (z; \p {\sf D})$.
\end{itemize}

\paragraph{Integral formulas.}
\begin{itemize}
\item
Cauchy's integral formula ($f$ is any smooth function):
$$
\frac{1}{2\pi i}\oint_{\p {\sf D}}
\frac{f(\zeta )d\zeta }{z-\zeta }
-\frac{1}{\pi}\int_{{\sf D}}
 \frac{\p_{\bar \zeta}f(\zeta )d^2 \zeta}{z-\zeta }
=\left \{\begin{array}{cl}
-f(z)\,, & \;\; z\in {\sf D} \\
0\,, &\;\; z\in {\bf C}\setminus {\sf D}
\end{array}\right.
$$
In particular,
$\oint_{\p {\sf D}}f(\zeta )d\zeta =2i
\int_{{\sf D}}\bar \p f (\zeta ) \, d^2 \zeta$.
\item
The Green formula:
$$
\int_{{\sf D}} f \Delta g d^2z = -\int_{{\sf D}}
 \nabla f \, \nabla g \, d^2z
+\oint_{\p {\sf D}}f \p_n g |dz|
$$
where the normal vector looks outward ${\sf D}$.
\item
The Dirichlet formula:
$$
u(z)=-\, \frac{1}{2\pi}\oint_{\gamma}
u(\zeta ) \p_{n} G(z,\zeta ) |d\zeta |
$$
for any function $u$
harmonic in ${\sf D^c} = {\bf C}\setminus {\sf D}$.
Here $G(z, \zeta )$ is the Green function
of the Dirichlet boundary
value problem in ${\sf D^c}$.
\end{itemize}

\subsection*{Appendix C}

Let us present some details of the $\hbar$-expansion
of the loop equation (\ref{loopeq1a}).
First of all we rewrite it in the form
$$
\frac{1}{2\pi}\int L(z, \zeta )
\lbracket \Delta \varphi (\zeta )\rbracket d^2 \zeta =
(\p \vphcl (z))^2 -
\lbracket \left ( \p ( \varphi (z) \! -\! \vphcl (z)\right )^2
\rbracket -
(2\! -\! \beta )\hbar  \lbracket \p^2 \varphi (z) \rbracket
$$
which is ready for the $\hbar$-expansion.
Here
$$
L(z, \zeta ) =\frac{ \p W(\zeta ) - \p \vphcl (z)}{\zeta -z}
$$
is the kernel of the integral operator in the l.h.s. (the
``loop operator"). The zeroth order in $\hbar$ gives
equation (\ref{large1}) which implies the familiar result
$\vphcl (z)=-\int_{{\sf D}} \log |z-\zeta |^2
\sigma (\zeta )d^2 \zeta$ for the $\vphcl$.
To proceed, one should insert the series
$$
\lbracket \varphi (z)\rbracket =
\vphcl (z) +\hbar \varphi_{1/2} (z) +\hbar^2 \varphi_1 (z)
+O(\hbar^3)
$$
(which corresponds to the $\hbar$-expansion (\ref{hbarexp})
of the free energy) into the loop equation and separate
terms of order $\hbar$, $\hbar^2$ etc. (In the notation
adopted in the main body of the paper $\hbar \varphi_{1/2}=
\varphi_{\hbar} +O(\hbar^2 )$.) The terms of order
$\hbar$ and $\hbar^2$ give:
$$
\begin{array}{l}
\displaystyle{
\frac{1}{2\pi}\int L(z, \zeta )
\lbracket \Delta \varphi_{1/2} (\zeta )\rbracket d^2 \zeta =
-(2 \! -\! \beta ) \p^2 \vphcl (z)}
\\ \\
\displaystyle{
\frac{1}{2\pi}\int L(z, \zeta )
\lbracket \Delta \varphi_1 (\zeta )\rbracket d^2 \zeta =
-\left [ \left ( \p \varphi_{1/2}(z)\right )^2 +
(2 \! -\! \beta ) \p^2 \varphi_{1/2} (z)\right ]
-\omega (z)}
\end{array}
$$
where
$$
\omega (z)= \lim_{\hbar \to 0} \left [ \hbar^{-2}
\lim_{z'\to z} \lbracket
\p \varphi (z) \, \varphi (z') \rbracket_c \right ]
$$
is the connected part of the pair correlator at merging
points. If the point $z$ is in ${\sf D^c}$,
then eq. (\ref{2trace1}) yields
$$
\lbracket
\p \varphi (z) \, \varphi (z') \rbracket_c
=2\beta \hbar^2 \p_{z}\p_{z'} \Bigl (
G(z, z') -\log |z-z'|\Bigr ) +O(\hbar^3 ))
$$
Since the r.h.s. is regular for all $z,z' \in {\sf D^c}$,
the points can be merged without any regularization and the
result does not depend on the particular limit $z' \to z$.
We thus obtain that the function $\omega (z)$ is proportional
to the Schwarzian derivative of the conformal map
$w(z)$:
$$
\omega (z)=\frac{\beta}{6}
\left ( \frac{w'''(z)}{w'(z)}- \frac{3}{2}
\left (\frac{w''(z)}{w'(z)}\right )^2 \right )
$$

The expansion of the loop equation can be continued
order by order. In principle, this gives
a recurrence procedure to determine the coefficients
$\varphi_k (z)$.
However, the equations of the chain are integral equations
in the plane, and it is not easy
to solve them explicitly.
Another difficulty is that
in general one can not extend these equations
to the interior of the support of eigenvalues
because the $\hbar$-expansion may break down
or change its form there.
Indeed, in the domain filled by the gas of eigenvalues
the microscopic structure of the gas becomes essential, and one
needs to know correlation functions at small scales.
Nevertheless, at least in the first two orders in $\hbar^2$
the equations above can be solved  assuming that
$z \in {\sf D^c}$. Note that in this region all
the functions $\varphi_k (z)$ are harmonic.
If these functions are known, the corresponding
expansion coefficients of the free energy in (\ref{hbarexp})
can be obtained
by ``integration" of the
variational formulas (\ref{var}).

The procedure of solving
the loop equation in the first
two orders in $\hbar$ is too technical
to be presented here. In the order $\hbar$ one is able
to find a complete solution which gives
formulas (\ref{phihbar}) and (\ref{F1/2})
mentioned in the main text. The solution in the next
order, $\hbar^2$, is much more difficult to obtain.
The results for
$\varphi_1$ and $F_1$ are still not available in full generality
(i.e., for general $\beta$ and $W$).
Nevertheless, for normal matrices with a general
potential ($\beta =1$) and with a connected support
of eigenvalues
the $F_1$-correction to the free energy can be found
explicitly by the method outlined above.
Here we
present the result (mostly for an illustrative purpose),
using the notation introduced in the main text (see
(\ref{chi}), (\ref{z(w)})):
$$
\begin{array}{lll}
F_1 & =& \displaystyle{
-\,\, \frac{1}{24 \pi} \oint_{|w|=1}
\Bigl (\log |z'(w)|\p_n \log |z'(w)| +
2\log |z'(w)| \Bigr )|dw|}
\\&& \\
&&\displaystyle{ - \,\, \frac{1}{24 \pi}\left [
\int_{{\sf D}} |\nabla \chi |^2 \, d^2 z + 2\oint_{\p {\sf D}}
\kappa \chi \, |dz | \right ]}
\\&&\\
&&+\,\, \displaystyle{\frac{1}{8\pi}
\left [\int_{{\sf D}} |\nabla \chi |^2 \, d^2 z -
\oint_{\p {\sf D}}
\chi \p_n \chi^H \, |dz | \right ]
-\frac{1}{16\pi} \int_{{\sf D}}\Delta \chi d^2 z
+c_0}
\end{array}
$$
where $c_0$ is a numerical constant
and $\kappa (z)=\p_n \log \left |
\frac{w(z)}{w'(z)}\right |$ is the local curvature of the boundary.
For quasiharmonic potentials only the first integral
survives.

\subsection*{Appendix D}

Here we demonstrate how
the variational technique works
with the 2-trace correlator (\ref{2trace}).
We shall use the variational formulas (\ref{var})
in the following equivalent version.
Set $\delta W(z)=\epsilon g(z)$, where $g$ is an arbitrary smooth
function and $\epsilon \to 0$. Then, in the first order
in $\epsilon$,
$$
\hbar \, \delta \lbracket \mbox{tr}\, f \rbracket
=\epsilon \lbracket \mbox{tr}\, f \, \mbox{tr}\, g\rbracket_c
$$
This relation allows one to find the connected
part of the two-trace correlation function by variation
of the known one-trace function.
Similar formulas hold for
variations of multi-trace functions.

We have, in the leading order in $\hbar$:
$$
\hbar \beta \delta \lbracket \mbox{tr}\, f \rbracket =
\delta\left (
\int_{{\sf D}} \sigma  f \, d^2 z \right )
=\int_{{\sf D}} \delta \sigma  f \, d^2 z +
\int_{\delta {\sf D}} \!\sigma  f \, d^2 z
\equiv I_1 + I_2
$$
The first integral, $I_1$, can be transformed
using the definition of $\sigma$ (\ref{sigma})
and the Green formula:
$$
I_1 =-\, \frac{1}{4\pi}
\int_{{\sf D}} \Delta (\delta W) f \, d^2 z
=\frac{\epsilon}{4\pi}
\int_{{\sf D}} \nabla g \, \nabla f \, d^2 z
-\, \frac{\epsilon}{4\pi}
\oint_{\p {\sf D}} f \p_n g \, |dz|
$$
The second integral is
$$
I_2 =-\, \frac{1}{4\pi}
\int_{{\sf D}} \Delta W (z) f(z) \delta n(z) |d z|
$$
where $\delta n(z)$ is to be taken from eq. (\ref{deltan}):
$\delta n(z)=\epsilon \p_n (g^H (z) \! -\! g(z))/\Delta W(z)$.
Summing the two contributions, we get (\ref{2trace}).

\subsection*{Appendix E}

In this Appendix we obtain the expansions of
the effective action ${\cal A}(z)$
$$
{\cal A}(z)=\frac{1}{2}|z|^2-\frac{1}{2}|\xi_0 |^2
-{\cal R}e
\int_{\xi_0}^{z} S(\zeta )\, d\zeta
$$
near the contour $\gamma =\p {\sf D}$.
We know that the first variation of ${\cal A}(z)$
vanishes on $\gamma$.
To find the second variation, we write
$$
2\delta {\cal A}(z) =
|\delta z|^2 -
{\cal R}e \,(S'(z)(\delta z)^2 ) \,,
\;\;\;\;z\in \gamma
$$
Let us represent $\delta z$
as a sum of normal and tangential
deviations w.r.t. the curve:
$\delta z =\delta_n z +\delta_t z$, then
$$
2{\cal R}e \,(S'(z)(\delta z)^2 )=
{\cal R}e \,(S'(z)(\delta_n z)^2 )+
{\cal R}e \,(S'(z)(\delta_t z)^2 )+
{\cal R}e \,(S'(z)\delta_n z \delta_t z)
$$
Using the obvious relations
$$
\frac{\delta_t z}{|\delta _t z|}=
\sqrt{\frac{\delta_t z}{\delta_t \bar z}}=
\frac{1}{\sqrt{S'(z)}}\,,
\;\;\;\;\;\;
\delta_n z =\mp i \left | \frac{\delta_n z}{\delta_t z}\right |
\delta_t z
$$
where the upper (lower) sign
should be taken for the outward (inward) deviation,
and the formula for the scalar product of 2-vectors
$\vec x, \vec y$ represented as complex numbers $x,y$,
$(\vec x, \vec y)={\cal R}e\, (x\bar y)$, we see that
the first and second terms in the r.h.s. are equal to
$-|\delta_n z|^2$ and
$|\delta_t z|^2$ respectively while the third one
vanishes since the vectors $\delta_n z$ and $\delta_t z$
are orthogonal.
Since
$|\delta z|^2=
|\delta_n z|^2 +|\delta_t z|^2$,
we obtain the desired result
$\delta {\cal A}(z)=
|\delta_n z|^2$.

The next terms of the expansion of the ${\cal A}(z)$
around the contour can be found in a similar way.
They are expressed through the curvature $\kappa$ and its
derivatives w.r.t. the arc length $s$ along the
curve. To perform the calculations
in next two orders we need the following
formulas for the $\kappa$
and $\kappa ' =d\kappa /ds$ through the
Schwarz function \cite{Davis}:
$$
\kappa (z)= \frac{i}{2}\frac{S''(z)}{(S'(z))^{3/2}}\,,
\;\;\;\;\;
\kappa '(z)=\frac{i}{2}\,
\frac{S'''(z)S'(z) -\frac{3}{2}(S''(z))^2}{(S'(z))^3}\,,
\;\;\;z\in \gamma
$$
For $z$ on the contour we have
$$
{\cal A}(z +\delta_n z)=|\delta_n z|^2 -
\frac{1}{6}{\cal R}e\, (S''(z)(\delta_n z)^3 )-
\frac{1}{24}{\cal R}e\, (S'''(z)(\delta_n z)^4 ) + \ldots
$$
Now, with the help of the formulas
for the curvature, it is easy to find that
$$
{\cal R}e\, (S''(z)(\delta_n z)^3 )=
2\kappa (z){\cal R}e \, \left (\frac{1}{i}
\left ( \sqrt{S'(z)}\delta_n z\right )^3 \right )
=2\kappa (z){\cal R}e \, \left (\frac{1}{i}
\left (\frac{\overline{\delta_t z}}{\delta_t z}\right )^{3/2}\!\!
(\delta_n z)^3 \right )
$$
whence
$$
{\cal R}e\, (S''(z)(\delta_n z)^3 )=
\pm 2\kappa (z)|\delta_n z |^3
$$
A similar computation gives
$
{\cal R}e\, (S'''(z)(\delta_n z)^4 )=
-6\kappa ^2 (z)|\delta_n z |^4
$
and we obtain the expansion (\ref{expanA}).
The expansion of $|w'(z)|$ around the contour
can be easily performed
with the help of the relations
$$
\kappa (z)=\p_n \log \left |
\frac{w(z)}{w'(z)}\right |,
\;\;\;\;\;
\p_n \log |w(z)| =|w'(z)|
$$
valid for $z\in \gamma$.

\end{document}